\documentclass[rsi, amsmath, amssymb, reprint, aip, longbibliography]{revtex4-1}

\usepackage{graphicx}% Include figure files
\usepackage{dcolumn}% Align table columns on decimal point
\usepackage{bm}% bold math
\usepackage[utf8]{inputenc}
\usepackage[T1]{fontenc}
\usepackage{mathptmx}
\usepackage{etoolbox}
\usepackage[colorlinks = true, linkcolor = blue, citecolor = blue, urlcolor = blue]{hyperref}% add hypertext capabilities

\newcommand{\aSi}{\mbox{a-Si}}
\newcommand{\mAumsq}{\mbox{mA/$\mu$m$^2$}}

\newcommand*{\citen}[1]{%
  \begingroup
    \romannumeral-`\x % remove space at the beginning of \setcitestyle
    \setcitestyle{numbers}%
    \cite{#1}%
  \endgroup   
}

\makeatletter
\def\@email#1#2{%
 \endgroup
 \patchcmd{\titleblock@produce}
  {\frontmatter@RRAPformat}
  {\frontmatter@RRAPformat{\produce@RRAP{*#1\href{mailto:#2}{#2}}}\frontmatter@RRAPformat}
  {}{}
}%
\makeatother
\begin{document}

\preprint{AIP/123-QED}

\title[Picosecond Josephson Samplers: Modeling and Measurements]{Picosecond Josephson Samplers: Modeling and Measurements}
\author{L. Howe}
    \affiliation{National Institute of Standards and Technology, Boulder CO 80305, USA}
    \affiliation{University of Colorado, Boulder CO 80309, USA}
    \email{logan.howe@nist.gov}
\author{B. van Zeghbroeck}%
    \affiliation{University of Colorado, Boulder CO 80309, USA}
\author{D. Olaya}
    \affiliation{National Institute of Standards and Technology, Boulder CO 80305, USA}
    \affiliation{University of Colorado, Boulder CO 80309, USA}
\author{J. Biesecker}
    \affiliation{National Institute of Standards and Technology, Boulder CO 80305, USA}
\author{C. J. Burroughs}
    \affiliation{National Institute of Standards and Technology, Boulder CO 80305, USA}
\author{S. P. Benz}
    \affiliation{National Institute of Standards and Technology, Boulder CO 80305, USA}
\author{P. F. Hopkins}
    \affiliation{National Institute of Standards and Technology, Boulder CO 80305, USA}

\date{\today}% It is always \today, today,

\begin{abstract}
Measurement of signals generated by superconducting Josephson junction (JJ) circuits require ultra-fast components located in close proximity to the generating circuitry. We report a detailed study of optimal design criteria for a JJ-based sampler which balances the highest sampler bandwidth (shortest \mbox{10\%--90\%} rise time) with minimal sampled waveform distortion. We explore the impacts on performance of a sampler, realized using a single underdamped JJ as the logical sampling element (\textit{the comparator}), due to the type of signal-comparator coupling scheme that is utilized (galvanic, inductive, or capacitive). In these simulations we emulate the entire waveform reconstruction sampling process, via comparator threshold detection, while sweeping the time location at which the waveform is being sampled. We extract the sampled waveform rise time (or FWHM) as a function of the comparator's Stewart-McCumber parameter and as a function of the coupling strength between the device under test and comparator. Based on our simulation results we design, fabricate, and characterize a cryocooled (3.6~K operating temperature) JJ sampler utilizing the NIST state-of-the-art Nb/amorphous-Si/Nb junctions. We separately sample a step signal and impulse generator co-located on-chip with the comparator and sampling strobe generator by implementing the same binary search comparator threshold detection technique during sampler operation as is used in simulation. With this technique the system is fully-digital and automated and operation of the fabricated device directly mirrors simulation. Our sampler technology shows a \mbox{10\%--90\%} rise time of 3.3~ps and the capability to measure transient pulse widths of 2.5~ps FWHM. A linear systems analysis of sampled waveforms indicate a 3~dB bandwidth of 225~GHz, but we demonstrate effective measurement of signals well above this -- as high as 600~GHz.
\end{abstract}

\maketitle

\section{\label{sec:intro}Introduction}
Circuits based on superconducting Josephson junctions (JJs) have wide applications ranging from signal metrology \cite{rufenacht2018impact, bauer2022josephson}, low-noise amplification \cite{castellanos2007widely, roy2018quantum, macklin2015near}, to novel computation (superconducting super- \cite{fourie2018digital, fourie2019coldflux} and quantum- computers \cite{blais2021circuit, krantz2019quantum, leonard2019digital, howe2022digital, castellanos2023coherence}) arising from their low power consumption, the nonlinear Josephson inductance, and area quantization of the single flux quantum (SFQ) pulses they can generate. Metrology of signal waveforms generated by JJ-based circuits is challenging due to the short duration of an SFQ pulse (a few picoseconds), and the fact the circuits are embedded in cryogenic systems -- precluding direct access using short cabling lengths. Despite these challenges, JJ-based samplers have successfully been used to diagnose operation of complex superconducting circuits via integration at key circuit nodes \cite{ketchen1985josephson}.

To minimize dispersion, and thus faithfully detect JJ-generated signals, an appropriate measurement technology must be located in close proximity to the superconducting circuits. An obvious choice is to incorporate a single JJ\cite{hamilton1979patent, wolf1985josephson,vanzeghbroeck1985model} or superconducting quantum interference device (SQUID) \cite{faris1979patent} as a logical comparator on the same chip with the target circuitry. This strategy is beneficial in terms of ease-of-integration -- the sampler can be co-fabricated on the same chip as the device under test (DUT) -- and simplicity of the experimental setup. Furthermore, this strategy provides a wide parameter space for the DUT-comparator coupling in terms of strength and type -- galvanic, inductive, or capacitive -- which can be easily defined and fixed during the design and fabrication phases.

Electro-optic (EO) sampling strategies \cite{kolner1986electrooptic, seitz2006characterization, wang1995optoelectronic} present another viable path to sampling waveforms generated by superconducting circuits and, due to the femtosecond timescale routinely available in pulsed laser systems today, can potentially offer rise times beyond that available to Nb-based superconducting samplers \cite{hegmann1995electro, keil1992electro}. However, a major limitation of EO sampler implementations arises from the need to incorporate an EO transduction element such as lithium- or niobium-tantalate. These transduction elements typically have large dielectric constants, e.g. $\epsilon_r = 43$ for LiTaO$_3$ \cite{kolner1986electrooptic}, and are frequently coupled to the electric field in an on-chip microwave transmission line, resulting in significant distortion of the target waveform (signal). Multiple reflections of the laser pulse from the substrate's top and bottom surfaces also confound the goal of low-distortion sampling \cite{seitz2006characterization, kolner1986electrooptic}. A final concern is the interplay between the EO sensitivity (or signal-to-noise ratio) and laser power, which has the potential for significant quasiparticle generation and degradation of the performance of the superconducting DUT circuitry. Nonetheless, EO sampling is a demonstrated technique for measurement of JJ-generated signals -- e.g. SFQ pulses \cite{wang1995optoelectronic}.

Initial JJ sampler efforts focused on Nb-based underdamped (latching) junctions, i.e. with Stewart-McCumber parameter $\beta_C = 2 \pi I_c C R^2 / \Phi_0 > 1$  \cite{wolf1985josephson, tuckerman1980josephson, faris1980generation, whiteley1988technologies, akoh1983direct, akoh1985real, sakai1983fluxon, harris1982electronically}, which still hold the highest bandwidth and shortest rise time of all clearly-demonstrated sampling techniques \cite{wolf1985josephson}. In the late 1980s, a commercial sampling oscilloscope and time domain reflectometer with advertised 70~GHz bandwidth and 5~ps risetime was developed based on this technology\cite{Whiteley1986patent, PPLComparison, whiteley1988technologies} -- however these bandwidths and rise times are insufficient for our applications targeting precision voltage metrology of circuits with rise times and pulse FHWM of $\lesssim 5$~ps. More recent efforts have been made using high temperature superconductors (HTSs) \cite{maruyama2005observation, hidaka1999high} in attempts to leverage the increased gap frequency \cite{suzuki2005progress} to yield a faster sampler response \cite{maruyama2007observation}. Because these realizations of HTS JJs were intrinsically shunted, and either critically- ($\beta_C = 1$) or over-damped ($\beta_C < 1$), a latching comparator could not be implemented. Instead, it was necessary to incorporate additional complexity in the form of SFQ-to-dc readout SQUIDs to determine comparator switching. As we will show in Sec.~\ref{sec:sim_results_step_imp}, shunting JJs into the critically-damped regime also increases their switching time and limits the bandwidth of such a sampler. Indeed, damping in HTS JJ samplers has served to significantly reduce the achieved bandwidth compared to expectations based simply on the increased gap frequency.

The ultimate goal of our sampler development lies in voltage and waveform metrology of JJ-based circuits generating signals with characteristic timescales of a few picoseconds, so the target sampling technology must have similar rise times and dynamic bandwidths of hundreds of gigahertz -- ideally faster and wider bandwidth than the target circuitry. To this end we pursue a latching JJ sampler co-located with the target-signal-generating circuits as this technology possesses the most straightforward, and flexible path to high-bandwidth, low-distortion sampling. Using NIST's state-of-the-art JJ process using Nb electrodes and amorphous-silicon (\aSi) barriers \cite{olaya2023josephson, olaya2023nb} we design a sampler prototype similar to [\citen{wolf1985josephson}] and demonstrate this strategy has sufficient speed for measuring circuits based on SFQ pulses.

\begin{figure}[t!]
    \centering
    \includegraphics[width=0.48\textwidth]{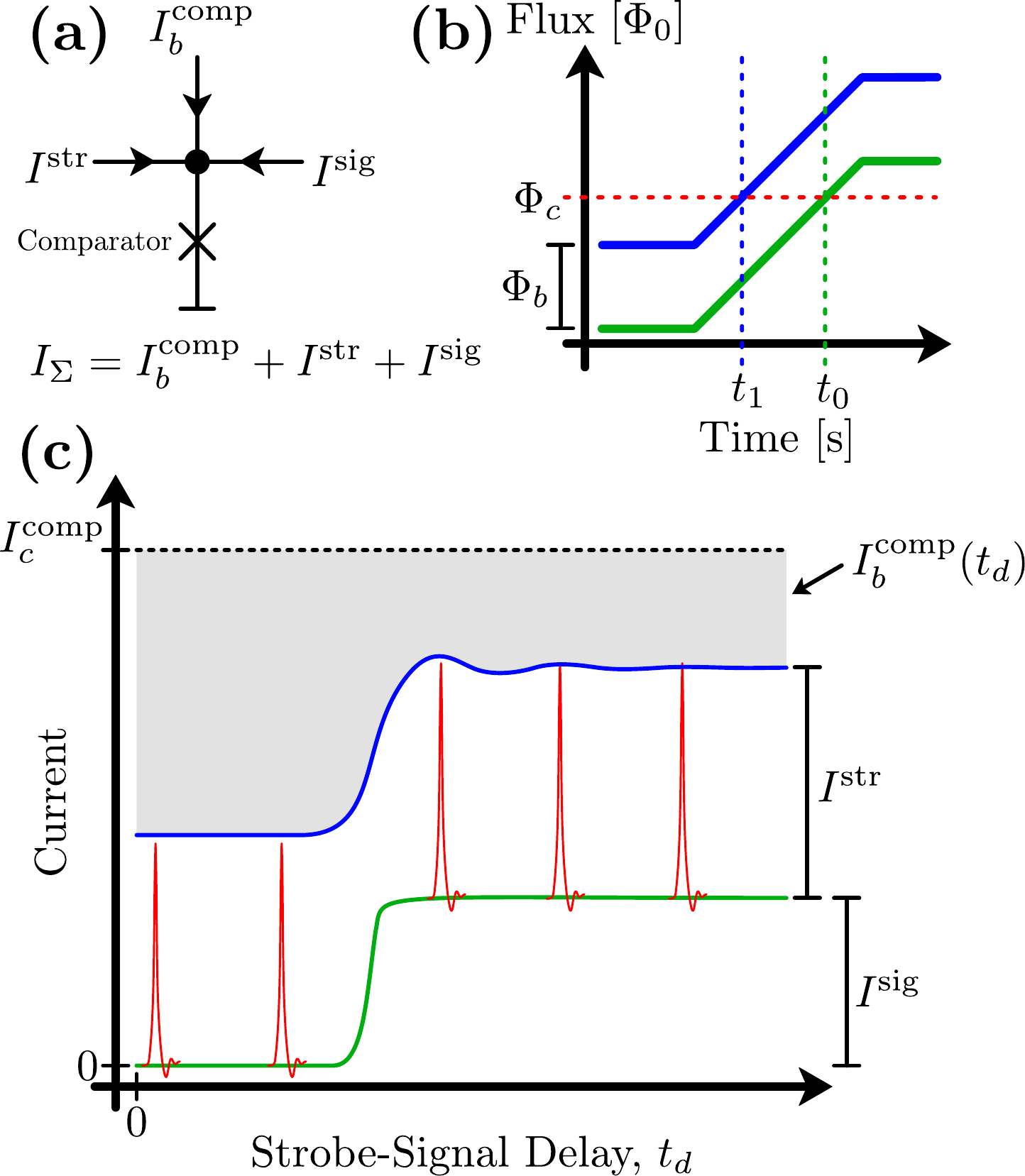}
    \caption{Simplified description of the latching JJ sampler operation. \textbf{(a)} Schematic showing how a large-critical-current JJ acts as a comparator. The comparator JJ is supplied with a dc bias current $I_b^\text{comp}$ from room temperature but also has current coming from an on-chip fast impulse generator $I^\text{str}$ (\textit{the sampling strobe}), and the signal current coming from the on-chip DUT, $I^\text{sig}$. Once the sum current, $I_\Sigma = I_b^\text{comp} + I^\text{str} + I^\text{sig}$ exceeds the comparator junction critical current, $I_c^\text{comp}$, it latches and generates an easily measurable dc voltage equal to the superconducting gap voltage. \textbf{(b)} Example of how a dc offset to a step-function magnetic flux-actuated strobe or DUT can be used to tune the relative delay between strobe emission and target waveform generation. For zero dc flux bias the strobe/DUT is triggered at time $t_0$ but when the dc flux bias is increased to $\Phi_b$ they are triggered at $t_1$. \textbf{(c)} Cartoon diagram showing the various currents during operation. The relative timing, a delay between the strobe (red pulses) and signal (green trace) is swept using the technique shown in (b), and the sum of the strobe and signal currents varies. Here we show the strobe moving relative to the signal but in practice we hold the strobe timing fixed and \textit{advance} the signal to traverse the strobe. At each delay $I_b^\text{comp}$ is tuned so $I_\Sigma$ always just exceeds $I_c^\text{comp}$ and is recorded. The $I_b^\text{comp}$ as a function of delay trace is thus the inverse (and offset) sampled signal (blue trace). Note that the sampled step signal (blue trace) is depicted with a shallower slope and with post-step oscillations to highlight possible artifacts resulting from the sampling measurement. }
    \label{fig:sampler_ball_and_stick}
\end{figure}

In this paper we present both simulations and test results of sampling the picosecond-duration output waveforms of two different DUTS with all JJs realized using underdamped (latching) Nb/\aSi/Nb JJs with a critical current density, $J_c$, of 0.22~\mAumsq. We demonstrate a novel technique for comparator threshold detection using a binary search that we apply in both simulation of the full sampling operation and measurement of fabricated devices. This allows us to fully simulate the entire sampling process in the same manner it is performed in the laboratory – for the first time ever in Josephson sampler technology – and enables strong agreement between our modeling and measurement results. In Sec.~\ref{sec:sampler_basics} we first review the operation of a latching Josephson sampler. In Sec.~\ref{sec:sampler_sims} we simulate a JJ sampler comprised of a single comparator with a critical current $I_c = 2$~mA, a pulse-generating \textit{strobe} circuit, and the circuit generating the target signal (the DUT). The first DUT that we consider outputs a step signal and the second outputs a finite-duration impulse. We implement these circuits in simulation with physically realizable components: a latching SQUID for the step signal DUT, and a Faris pulser \cite{faris1980generation} for the impulse generator DUT \cite{wolf1985josephson}. We will show there exists an interplay between the intrinsic sampler rise time / bandwidth and the dynamics of the DUT used to characterize it. In Sec.~\ref{sec:exp_results} we describe our experimental setup and present cryogenic measurements of sampler devices operated at 3.6~K for both the step function and impulse DUTs. Sec.~\ref{sec:bw_analysis} details a first-order linear systems analysis to extract an approximate sampler bandwidth. Additional simulation and measurement details are included in the Appendices.

\section{\label{sec:sampler_basics}Latching Josephson Sampler Operation}
The principle of operation of a latching Josephson sampler is shown in Fig.~\ref{fig:sampler_ball_and_stick}. A JJ with a large critical current, $I_c^\text{comp}$ (the \textit{comparator}) is dc-biased with current $I_b^\text{comp}$. The comparator is attached to an on-chip sampling pulse generator, the \textit{strobe}, and a number of DUT circuits. The comparator and the strobe generator make up the \textit{sampler}; where the strobe is used to selectively drive the total comparator current above $I_c^\text{comp}$ at a specified time. In general, only a single DUT circuit is operated and sampled at a time. When the sum ($I_\Sigma$) of the dc bias ($I_b^\text{comp}$), the strobe current pulse ($I^\text{str}$), and the DUT signal ($I^\text{sig}$) exceeds the comparator $I_c^\text{comp}$ (the \textit{threshold}), the comparator voltage rapidly latches to the superconducting gap voltage. In this way, the entire DUT signal waveform can be sampled by changing the delay of the strobe relative to the DUT signal, as illustrated in Fig~\ref{fig:sampler_ball_and_stick}. The comparator remains latched at the Nb gap voltage of 2.8~mV until $I_\Sigma$ is dropped below the comparator return current -- making it an experimentally straightforward task to both determine whether the comparator has latched and to reset it for the next sampling event. We define the interval between setting and resetting all biases, during which the strobe and DUT are triggered and the comparator is allowed to latch (depending on $I_\Sigma$), as a \textit{sampling period}. Due to this required reset, the dc biases are actually square pulses.

\begin{figure}[t!]
    \centering
    \includegraphics[width = .48\textwidth]{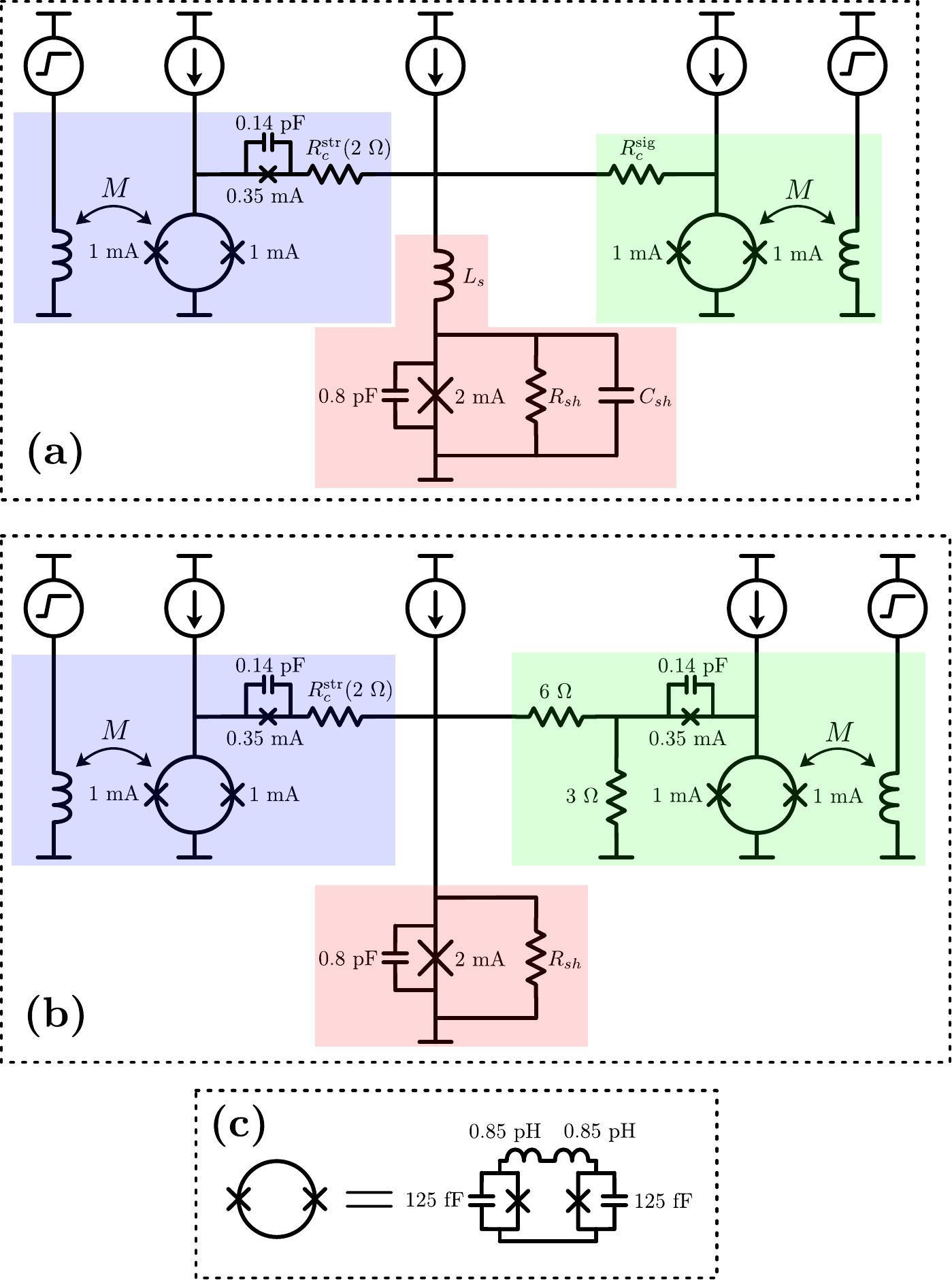}
    \caption{ Schematics of the circuits used to perform simulations, where \textbf{(a)} implements a latching SQUID step signal DUT and \textbf{(b)} uses a Faris pulser impulse DUT. The comparator and additional reactive coupling elements (Sec.~\ref{sec:reactive_coupling_summary} and Appendix~\ref{app:reactive_coupling}) are highlighted in red, the Faris pulser strobe generator in blue, and the DUT in green. Each Faris pulser is comprised of a symmetric SQUID with 1~mA $I_c$ SQUID JJs and a 0.35~mA $I_c$ output JJ. The sampler is comprised of the blue (strobe) and red regions with the strobe coupled to the comparator via $R_c^\text{str}$ whose nominal value is $2~\Omega$. We apply static current biases -- corresponding to 90\% of the subcircuit $I_c$ -- to the strobe, comparator and DUT, and magnetically actuate the strobe and DUT using step current pulses. $M$ represents the mutual inductance between the flux trigger line and the SQUID loops. The strobe pulse time is held fixed and we sweep the DUT signal across the strobe pulse by varying the relative delay of the DUT trigger. All static component values are listed numerically, while all swept parameters are indicated symbolically. $R_{sh}$ is used to explicitly tune the comparator $\beta_C$, while $R_c^{\text{sig}}$ tunes the coupling strength of the SQUID DUT. In Appendix~\ref{app:reactive_coupling} we discuss the sampler response under reactive coupling by sweeping $C_{sh}$ and $L_s$. \textbf{(c)} Circuit values for the symmetric SQUIDs.}
    \label{fig:sim_schematic}
\end{figure}

Latching JJ samplers have historically operated with sampling periods on the order of a few microseconds inside a slower feedback loop (with integration times \mbox{0.1~s--1~s}) configured to maintain the total current into the comparator exactly at $I_c^\text{comp}$ on slow timescales\cite{wolf1985josephson, tuckerman1980josephson, whiteley1988technologies, akoh1983direct, askerzade2006josephson}. A secondary bias current is typically applied at the slower timescale which sweeps the relative delay between triggering the strobe and DUT by causing one to generate its waveform slightly earlier on the rising edge of the flux pulse. Thus the sweep bias causes the strobe and signal to traverse one another at the slower integrator feedback timescale, resulting in full sampling of the target waveform at the sweep generator frequency.

Note that the purpose of the integrator feedback is to determine the change in bias required to latch the comparator at a given relative timing delay, shown as the gray $I_b^\text{comp}(t_d)$ region in Fig.~\ref{fig:sampler_ball_and_stick}, due to the value of the target signal when the strobe arrives. %A modern implementation may be realized by omitting
In this work, we omitted the sweep bias and feedback loop in favor of a digitizer and two programmable current sources. One current source is used to first set a fixed delay by setting a small added current bias to the DUT flux trigger. Then the digitizer timestream mean voltage -- windowed over a portion of a single sampling period after the strobe is known to have triggered -- can be used as a simple metric to determine if the comparator has latched. The second current source, supplying $I_b^\text{comp}(t_d)$, is dithered to determine the (e.g. 50\%) latching threshold for each programmed delay. In practice we implement a binary search algorithm to determine $I_b^\text{comp}(t_d)$. This experimental procedure is deliberately chosen to mirror the simulation strategy described in the following section to permit us to much more accurately simulate experimentally-realized samplers \cite{vanzeghbroeck2023josephson}.

\begin{figure}[t!]
    \centering
    \includegraphics[width = .48\textwidth]{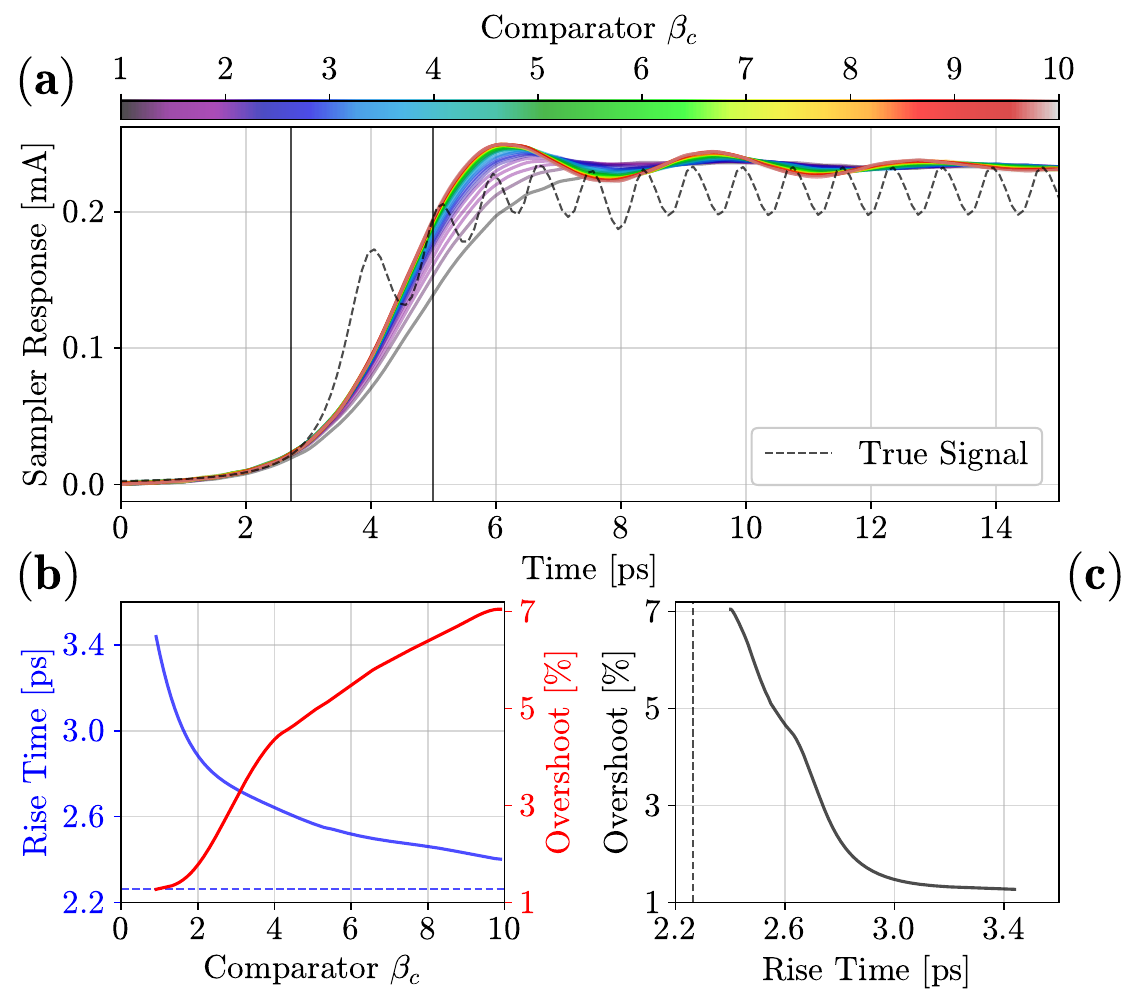}
    \caption{ Simulation of the sampler response with a step signal while varying the comparator $\beta_C$. Here $\beta_C$ is tuned by sweeping the comparator's explicit shunt resistance $R_{sh}$. \textbf{(a)} Solid colored lines show the sampler output of the latching SQUID for various $\beta_C$. The black dashed line shows the simulated SQUID signal at its output node which the sampler is attempting to reconstruct. Vertical black lines show the 10\% and 90\% times for the SQUID signal. \textbf{(b)} The blue (red) line shows the sampler rise time $t_r$ (overshoot) as a function of $\beta_C$. The blue dashed line shows the SQUID rise time, i.e. the minimum measurable $t_r$. \textbf{(c)} Illustration of the tradeoff between minimal sampled signal distortion (i.e. overshoot) and rise time. The SQUID rise time is denoted here by the black dashed line.}
    \label{fig:tres_vs_Rshunt}
\end{figure}

\begin{figure}[t]
    \centering
    \includegraphics[width = .48\textwidth]{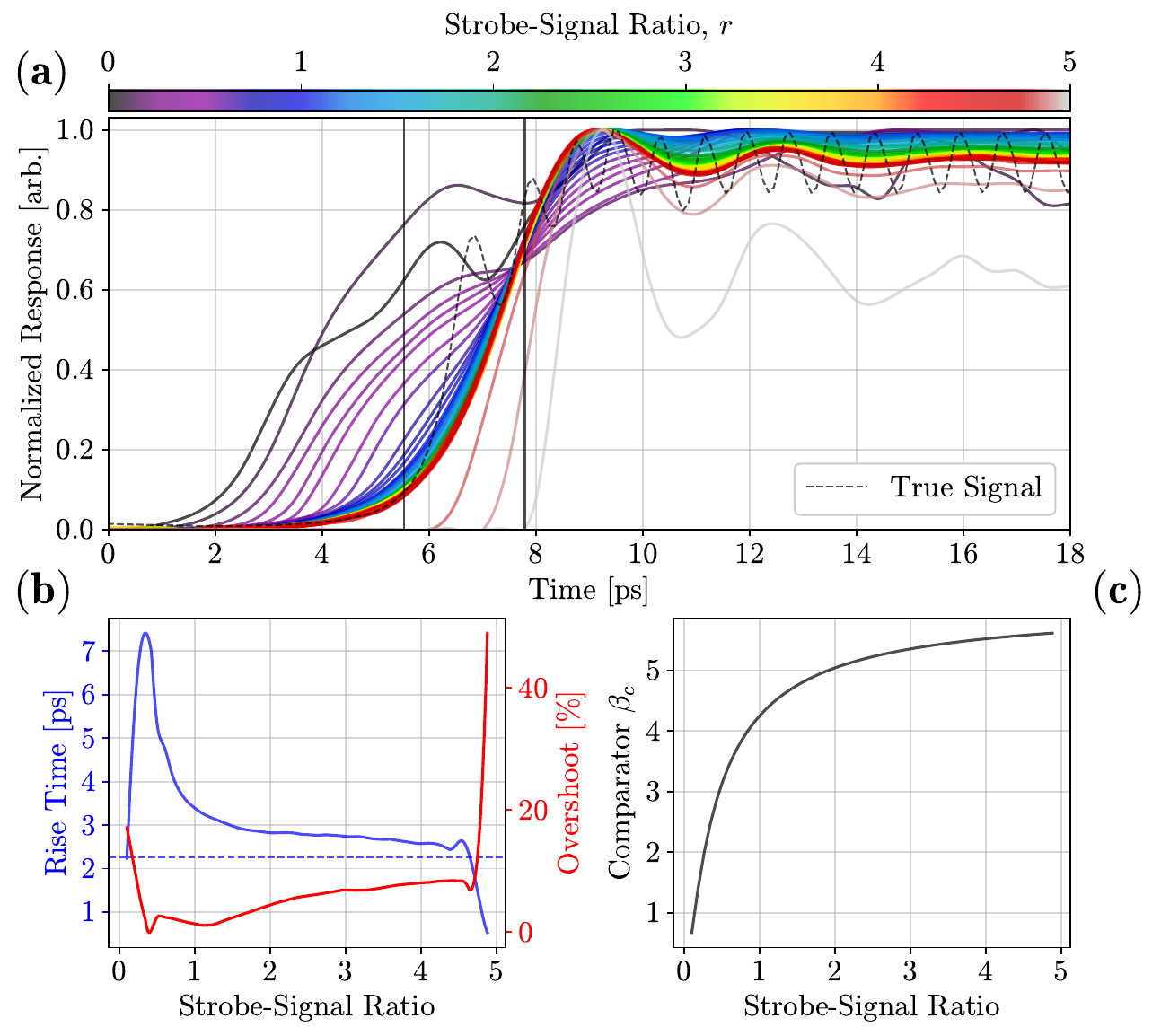}
    \caption{ Simulation of the sampler step signal response while varying the strobe-to-signal ratio ($r$) by tuning the signal coupling resistance, $R_c^\text{sig}$, linearly from 0.5~$\Omega$ to 25~$\Omega$ keeping the strobe coupling fixed at $R_c^\text{str} = 2~\Omega$. \textbf{(a)} Solid colored lines show the normalized sampled SQUID signal for various $r$. The black dashed line and vertical black lines show the simulated SQUID signal at its output node, and the 10\% and 90\% rise times, respectively, as in Fig.~\ref{fig:tres_vs_Rshunt} \textbf{(b)} The blue (red) line shows the sampler rise time $t_r$ (overshoot) as a function of $r$. The blue dashed line shows the SQUID rise time, i.e. the minimum measurable $t_r$. Large excursions in $t_r$ and overshoot, resulting from significant sampled waveform distortion, are observed for extreme values of $r$. For $r \lesssim 1$ the strobe does not strongly determine which part of the signal to sample, and for $r \gtrsim 4.5$ the sampler response becomes a strong convolution of not just the comparator dynamics and the target waveform, but also of the strobe and its post-pulse behavior. \textbf{(c)} The comparator $\beta_C$ as a function of $r$. $\beta_C$ is dominated by the $R_c^\text{str} = 2~\Omega$ resistor, but for $R_c^\text{sig} \lesssim 4~\Omega$ the comparator $\beta_C$ is appreciably affected by lowering the signal coupling resistance. As expected, when $\beta_C \sim 1$ the sampled response is heavily distorted due to unstable comparator latching. }
    \label{fig:tres_vs_Rcouple}
\end{figure}

\section{\label{sec:sampler_sims}Sampler Performance Simulations}
Simulations of the ideal and ultimate JJ sampler performance using a delta function strobe and zero-rise-time step signal have recently been performed \cite{vanzeghbroeck2023josephson, vanzeghbroeck2024switching} so here we focus on physically realizable circuits and signals. To extract rise time we simulate the circuit shown in Fig.~\ref{fig:sim_schematic}(a) -- where the DUT is a latching SQUID. To simulate sampling of a finite-duration pulse we use the circuit shown in Fig.~\ref{fig:sim_schematic}(b). For a step signal the important metrics are the $\mbox{10\%--90\%}$ rise time -- defined as the \textit{rise time}, $t_r$ -- and the distortion of the sampled signal -- which is quantified as the \textit{overshoot} during the initial rise in sampler signal relative to the asymptotic, or steady-state, sampler response after the SQUID has latched \cite{aastrom2007feedback}. When considering the response for imaging an impulse we consider the full width at half maximum (FWHM) and the \textit{undershoot} following the main peak to benchmark performance. In both cases there is an effective \textit{speed limit} imposed by the intrinsic dynamics of the target signal generators beyond which the sampler cannot be characterized. For the step signal DUT this is the \mbox{10--90\%} rise time of the SQUID itself, and for the impulse generator DUT this is the FWHM of the Faris pulser pulse directly at its output. These effects are due to the fact that the sampler output is the convolution of the comparator response and the DUT signal.

We use the WRspice circuit simulation package %\footnote{Commercial instruments and software are identified in this paper in order to adequately specify the experimental procedure. Such identification does not imply recommendation or endorsement by NIST, nor does it imply that the product identified is necessarily the best available for the purpose.}%
\cite{NISTdisclaimer, wrspice} to perform our simulations and use the built-in resistive- and capacitively-shunted junction (RCSJ) model. We choose the WRspice quasiparticle model giving a quasiparticle contribution with an exponential dependence on the bias current \cite{wrspice, xic, fourie2018digital}. We use JJ parameters characteristic to our Nb/\aSi/Nb fabrication process \cite{olaya2019planarized} (Appendix~\ref{app:fab}): a specific capacitance of 80~fF/$\mu$m$^2$, characteristic voltage of $I_cR_n = 1.5$~mV, and a subgap resistance of 60~$\Omega$. We remove the implicit JJ shunts in WRspice and provide explicit shunts in the circuit definition when needed. To determine the threshold comparator bias at each strobe-signal delay we implement a binary search technique in which only single sampling periods at a time are simulated and comparator latching is determined by a simple voltage threshold in the absence of noise (see Appendix~\ref{app:binary_search}).

\subsection{\label{sec:sim_results_step_imp}Step Signal Response}

First we consider the sampler dynamics when sampling the step signal created by the latching SQUID, shown in Fig.~\ref{fig:sim_schematic}(a), while various circuit parameters are swept. As previously mentioned, the goal of this study is to determine optimal design considerations for a JJ sampler -- specifically in terms of an ideal target for the comparator Stewart-McCumber parameter in the presence of an external coupling network $\beta_C = 2 \pi I_c C_{sh,eq} R_{sh,eq}^2 / \Phi_0$, and the most favorable signal coupling method: galvanic, capacitive, or inductive. Here $C_{sh,eq}$ and $R_{sh,eq}$ are the equivalent capacitance and resistance shunting the comparator due to the addition of the connected circuits in Fig.~\ref{fig:sim_schematic}. First we perform a $\beta_C$ sweep, by tuning the value of the explicit comparator shunt resistor $R_{sh}$, and extract $t_r$ and the overshoot from the sampled output waveform. The comparator $\beta_C$ is $R_{sh, eq} = [R_{sh}^{-1} + (R_c^\text{str})^{-1} + (R_c^\text{sig})^{-1}]^{-1}$, and is thus effectively bounded above by $R_c^\text{str} = 2~\Omega$ for our case. Results of this simulation sweep are shown in Fig.~\ref{fig:tres_vs_Rshunt} and reveal a clear tradeoff between minimal distortion sampling of the SQUID signal and the rise time. Indeed, low-distortion sampling -- corresponding to an overshoot of $<1\%$ -- requires a nearly critically-damped comparator. Given the weaker relation between $\beta_C$ and $t_r$ it is more favorable to enforce $\beta_C \lesssim 3$, as this is effectively the ``knee" in Fig.~\ref{fig:tres_vs_Rshunt}(b,c) above which small gains in rise time result in significant distortion of the signal when sampled.

The next obvious question in Josephson sampler design is how large the relative amplitudes of the strobe and signal should be. Previous Josephson sampler implementations\cite{faris1980generation, maruyama2007observation,tuckerman1980josephson}  have operated with strobe-to-signal ratios, $r = I^\text{str} / I^\text{sig}$, from unity to over 10 -- with the most successful sampler\cite{wolf1985josephson} possessing two DUTs with strobe-to-signal ratios of $\sim 1.8$ and $\sim 3$. To determine the sampler response as a function of the strobe-to-signal ratio we choose the coupling resistor $R_c^\text{sig}$ as our free parameter and conduct a rise time simulation, as detailed above, and with $R_{sh} = 2~\Omega$ (maximum $\beta_C = 5.6$). For these simulations, because we sweep $R_c^\text{sig}$ over an order of magnitude, the amount of SQUID bias current which branches upon latching of the SQUID and is then injected into the comparator is significant. This reduces the amplitude of the signal current relative to the full range of comparator bias needed to ensure $I_{min}$ never latches the comparator while $I_{max}$ always latches. To compensate for these effects we peak-normalize the sampler output for each $R_c^\text{sig}$ trace.

\begin{figure}[b!]
    \centering
    \includegraphics[width = .48\textwidth]{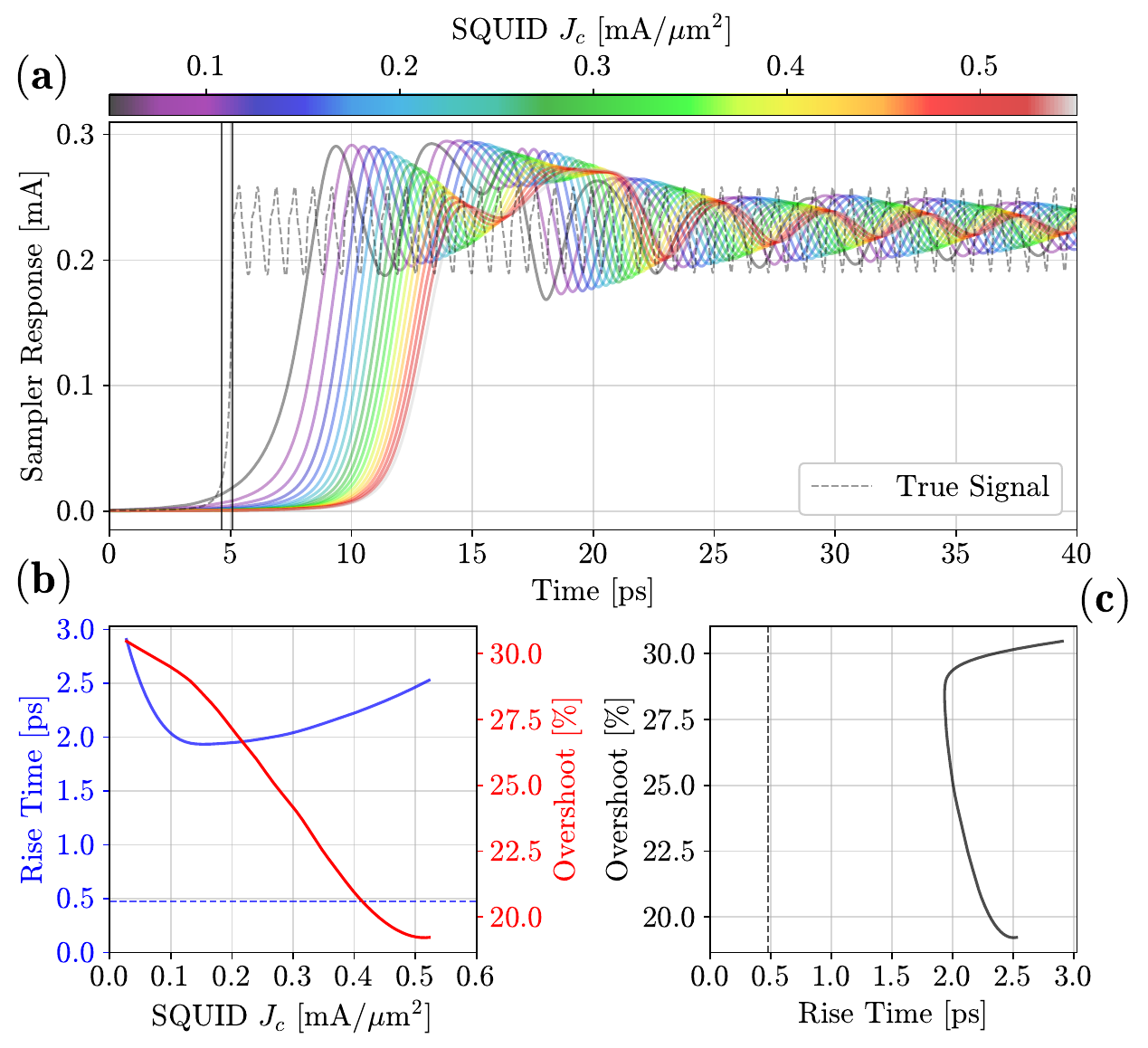}
    \caption{ Multi-$J_c$ process sampler simulation results. We increase the SQUID latching speed by reducing $L_{SQ}$ to 0.19~pH and then sweep the $J_c$ of only the SQUID JJs. The comparator Stewart-McCumber parameter is held at $\beta_c = 10$. \textbf{(a)} Solid colored lines show the sampled SQUID signal as a function of the SQUID $J_c$. The black dashed trace shows the simulated SQUID signal at its output node for the highest $J_c$ of 0.5~mA/$\mu$m$^2$. Vertical black lines show the 10\% and 90\% times for the SQUID signal. \textbf{(b)} The blue (red) line shows the sampler rise time $t_r$ (overshoot) as a function of $J_c$. The blue dashed line shows the SQUID rise time, i.e. the minimum measurable $t_r$. \textbf{(c)} Signal distortion (overshoot) as a function of $t_r$. The SQUID rise time is denoted here by the black dashed line.}
    \label{fig:high_Jc_squid_sim}
\end{figure}

Fig.~\ref{fig:tres_vs_Rcouple}  shows the simulated sampler output while varying $r$. Significant signal distortion and degradation in the rise time is evident in Fig.~\ref{fig:tres_vs_Rcouple}(a) for $r \lesssim 1$. Additionally, for $r \gtrsim 4.75$ we see that the sampler is unable to reconstruct the target signal with appreciable fidelity. This is due to the finite, non-zero tail produced by Faris pulsers, i.e. the strobe, implemented with large sub-gap resistance JJs \cite{cui2017scanning}. When the relative height of the strobe's post-pulse tail is approximately equal to the SQUID DUT signal amplitude the sampler instead attempts to measure both the target waveform and the strobe pulse tail simultaneously. Our cutoff of $r \sim 4.75$ indicates the post-pulse tail amplitude is approximately 1/5th the peak pulse height, which is in good agreement with the strobe Faris pulser output seen in simulation. Clearly, for our implementation, the strobe-signal ratio has an optimum value of 2--3 as this optimizes the rise time while allowing variation in $r$ without compromising performance.

\subsubsection{\label{sec:high_Jc_sim}High Current Density JJs: Faster DUTs and Sampler}
The extracted sampler rise times in the simulations in Sec.~\ref{sec:sim_results_step_imp} are a result of the convolution of the sampler response and the finite-width of the SQUID DUT step-function output\cite{maruyama2007observation}. The sampler response is limited by the comparator dynamics and the finite-width of the strobe pulse; both can be improved by increasing the ratio of the critical current to junction capacitance, as shown elsewhere\cite{vanzeghbroeck2023josephson,vanzeghbroeck2024switching,vanzeghbroeck1985model, harris1979turnon,WhiteleyTJM}. The rise time of the SQUID DUT output is limited by the SQUID $\beta_c$ \cite{harris1982electronically, vanzeghbroeck2024switching} and both the $L_{SQ}/R$ and $RC$ rise times. The relevant resistances for $R$ are both the SQUID normal resistance and the DUT-comparator coupler $R_c^\text{sig}$.  

Here we explore the feasibility of creating a real circuit (DUT) which could be used to measure the true ultimate rise time of the comparator. Namely we explore avenues to increase the SQUID latching speed, while making no changes to the comparator+strobe (sampler) circuit, and test whether we can observe improved sampled rise times compared to those presented in Sec.~\ref{sec:sim_results_step_imp}. Decreasing the SQUID self-inductance, $L_{SQ}$, and JJ capacitance yields a significantly faster SQUID response. In our simulations we can reduce the SQUID switching time by over a factor of 4.5, resulting in a \mbox{10\%--90\%} rise time of 0.47~ps, by lowering $L_{SQ}$ to 0.19~pH and the SQUID JJ capacitance to 50~fF. These values are a reduction by a factor of 10 and 2.5 relative to $L_{SQ}$ and $C_{JJ}$ used in Sec.~\ref{sec:sim_results_step_imp}, respectively. Both reductions are physically possible by (a) moving to a geometry where the SQUID loop axis is in the plane of the substrate rather than orthogonal \cite{wolf1985josephson}, and (b) varying the JJ process critical current density \cite{olaya2023josephson} at fixed specific capacitance (80~fF/$\mu$m$^2$). 

We simulated the sampled rise time of this faster SQUID DUT, fixing $L_{SQ}$ at 0.19~pH and sweeping the SQUID $J_c$ while the sampler circuit remained unchanged ($J_c$ held fixed at 0.2~\mAumsq). In WRspice\cite{NISTdisclaimer} this is accomplished by fixing $I_c$ and sweeping the JJ self-capacitance -- i.e. a JJ fabricated in a higher $J_c$ process will have a smaller area for a fixed $I_c$ and thus a smaller capacitance. Fig.~\ref{fig:high_Jc_squid_sim} shows the results of this ``multi-$J_c$" device with a comparator $\beta_c$ of 10 (fastest sampler response of Fig.~\ref{fig:tres_vs_Rshunt}). We observe from these results that, even with the SQUID $J_c$ a factor of two below our original design point of 0.2~\mAumsq, the reduction in $L_{SQ}$ still allows us to \textit{measure} and \textit{demonstrate} a sampler $t_r$ which is faster than that obtained in the simulations of Sec.~\ref{sec:sim_results_step_imp}; note however, that without changing the sampler circuit parameters, the minimum measured rise time is $\sim 1.9$~ps even with a SQUID rise time of $\sim 0.5$~ps. Thus we may interpret these results as revealing the intrinsic sampler rise time is approximately 1.9~ps

The dependence of the minimum quantifiable sampler rise time on the SQUID $J_c$ is nontrivial and is actually deleterious once the SQUID $J_c$ exceeds $\sim 0.3$~\mAumsq. I.e., the most fruitful path to produce a DUT whose rise time is less than the intrinsic comparator capability, and guarantees we may fully explore this contribution to the sampler's $t_r$, is to reduce the SQUID self-inductance. Of course, care must be taken to produce DUTs whose response is fast enough that the ultimate limiting behavior of the sampler (and not the DUT-sampler system convolution) may be faithfully characterized. Co-fabricating a DUT with a higher $J_c$ than that of the comparator is purely a technique for extracting the intrinsic sampler rise time and bandwidth, at a cost of more processing steps, and is not of large practical import.

\begin{figure}[t!]
    \centering
    \includegraphics[width = .48\textwidth]{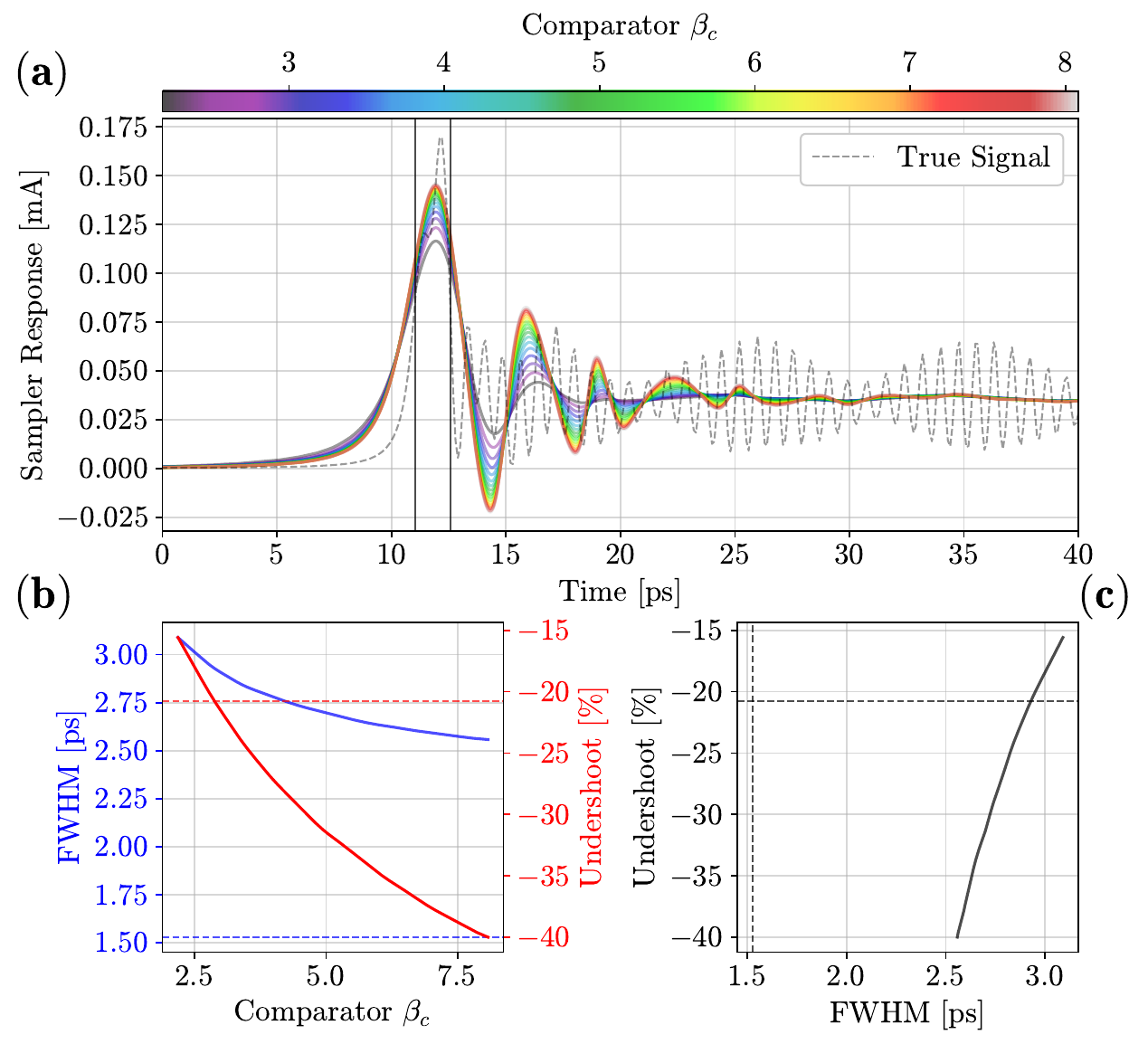}
    \caption{ Sampler response simulation versus comparator $\beta_C$ while sampling the output of a Faris pulser DUT coupled to the comparator via a 6~$\Omega$ series, 3~$\Omega$ to ground current divider. \textbf{(a)} Sampled Faris pulser response as a function of $\beta_C$ (solid colored lines). The dashed black line shows the ideal pulse while the vertical black lines show the half-maxima values. \textbf{(b)} Sampled pulse FWHM (blue) and undershoot (red). The undershoot is calculated as the largest excursion after the main pulse below the asymptotic response -- which is then normalized by the primary peak height. Dashed lines indicate the values intrinsic to the ideal Faris pulser signal. \textbf{(c)} Relation of the sampled pulse FWHM and its undershoot. The dashed vertical (horizontal) line shows the intrinsic pulse FWHM (undershoot).}
    \label{fig:faris_pulser_pulse_vs_Rshunt}
\end{figure}

\subsection{\label{sec:faris_pulser_sim}Finite Impulse Response}
To conclude our design studies of latching JJ samplers we focus on the sampler's finite impulse response -- in this case realized using a Faris pulser identical to the strobe generator and coupled to the comparator using a current divider comprised of a 6~$\Omega$ series resistance, and 3~$\Omega$ parallel resistance to ground. The current divider is chosen to make $\beta_C$ for the Faris pulser DUT and strobe approximately the same while enforcing a strobe-signal ratio of $\sim 2$. Here we perform a single comparator $\beta_C$ sweep and characterize the reconstructed waveform FWHM and undershoot. Fig.~\ref{fig:faris_pulser_pulse_vs_Rshunt} shows a much higher sensitivity to signal distortion at the sampler output relative to the SQUID DUT step signal response case. While the sampled Faris pulser pulse FWHM is nearly static as $\beta_C$ is tuned, significant improvement in distortion is achieved when the comparator is close to critically damped. 

\subsection{Reactive Coupling\label{sec:reactive_coupling_summary}}
Leveraging the learnings of Sec.~\ref{sec:sim_results_step_imp} we conclude that an optimally flexible JJ sampler, capable of reconstructing the fastest signals with minimal distortion is achieved when $\beta_C \sim 1$--4 and with a strobe-signal ratio of 2--3. However, this is in the case of \textit{galvanically}-coupled DUTs. It is certainly possible to \textit{reactively}-couple the DUTs using either capacitors in parallel or inductors in series with the comparator ($C_{sh}$ and $L_s$ in Fig.~\ref{fig:sim_schematic}(a)). Such a coupling scheme is attractive in scenarios where the sampler and DUTs are housed on separate chips -- such as a scanning SQUID sampler \cite{cui2017scanning} or in the case of a flip-chip bonded sampler-DUT multi-chip-module. However, bounds on the magnitudes of the reactive couplers need to be established from a perspective of shortest rise time/ highest bandwidth and minimum signal distortion. Intuition suggests that the couplers should behave as small perturbations to the comparator -- i.e. by only weakly modifying its $\beta_C$, $RC$ time, or $L/R$ time. Indeed, the results of our simulations (detailed in Appendix~\ref{app:reactive_coupling}) place upper bounds on the coupler capacitance and inductance of $\sim 1$~pF and $\sim 1$~nH, respectively, which corresponds almost exactly to the values of the comparator JJ self-capacitance and its Josephson inductance. Despite these relatively small limits on coupler reactance, sufficient signal coupling is possible at these levels due to the speed of the signals of interest.

\section{\label{sec:exp_results}Experimental Results}

\begin{figure*}
    \centering
    \includegraphics[width = .95\textwidth]{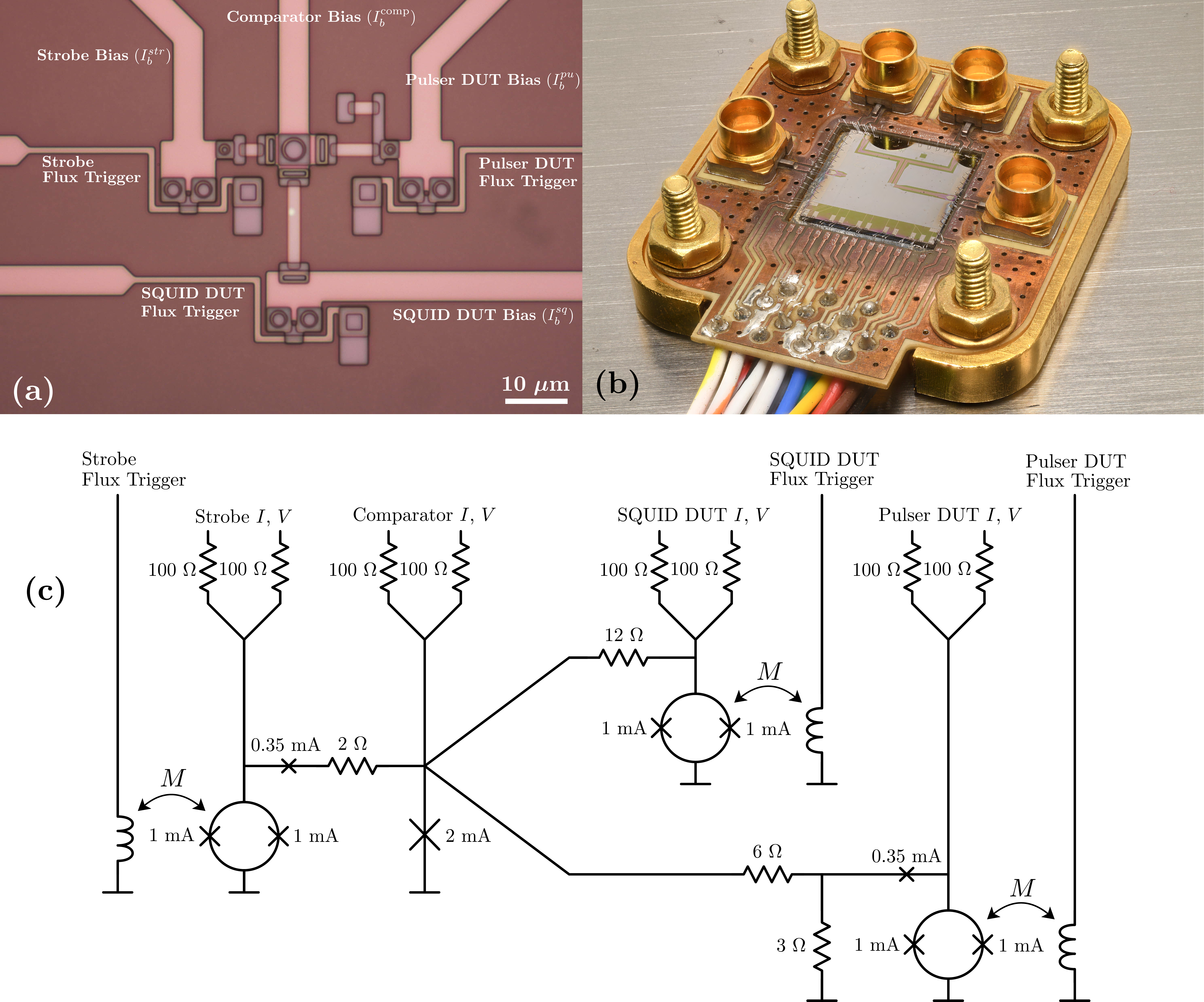}
    \caption{ \textbf{(a)} Micrograph of the sampler with dc bias and high-speed flux trigger lines denoted. Photo is taken immediately prior to lithography and deposition of the final insulator layer and sky plane. \textbf{(b)} Photo of a packaged (copper box lid removed) sampler device die showing the high-speed flux trigger SMP connections, dc signal lines, and wirebond interconnects. The \mbox{1~cm$~\times~$1~cm} sampler chip is PdAu backside-metallized and InSn-soldered directly to the package's copper bottom. \textbf{(c)} Electrical schematic of the sampler. Each dc bias lead and voltage tap have an on-chip 100~$\Omega$ bias and thermalization resistor to mitigate thermal current noise. The three flux trigger lines, brought in on high-bandwidth coaxial cables (20~GHz) and SMP connectors, are thermalized with 10~dB attenuators on the cryostat's 3~K stage. These flux triggers are supplied using step function voltage pulses from a multichannel, high-speed AWG (65 GSa/s, 25 GHz analog bandwidth).}
    \label{fig:sampler_pic_and_fabricated_device_schematic}
\end{figure*}

\subsection{\label{sec:fab}Device Design, Fabrication, and Packaging}
As a launching point for developing a flexible JJ sampler technology for waveform metrology and diagnostics of SFQ digital logic circuits, we fabricated a sampler using unshunted amorphous silicon barrier \mbox{(Nb/\aSi/Nb)} JJs with a critical current density of $J_c = 0.22$~\mAumsq. The fabricated circuit is shown in Fig~\ref{fig:sampler_pic_and_fabricated_device_schematic}(a) and the packaged \mbox{1~cm$~\times~$1~cm} chip is shown in Fig~\ref{fig:sampler_pic_and_fabricated_device_schematic}(b). We select this junction technology due to its potential to scale to significantly higher $J_c$, a critical parameter in increasing latching junction operating speed \cite{harris1979turnon, vanzeghbroeck2024switching}, when compared to more common \mbox{Nb/Al-AlO$_x$/Nb} JJs \cite{olaya2023josephson,tolpygo2017properties}.

The chosen topology is shown in Fig~\ref{fig:sampler_pic_and_fabricated_device_schematic}(c) and is similar to [\citen{wolf1985josephson}]. The sampler circuit is comprised of four primary components: the comparator ($I_c^\text{comp} = 2.21$~mA), a strobe pulse generator, a step signal DUT, and an impulse generator DUT. The strobe pulse and impulse generator DUT are realized using Faris pulsers \cite{faris1980generation}, consisting of a two-junction, symmetric dc SQUID (single junction $I_c = 1.11$~mA), and a smaller output pulser junction with $I_c = 385~\mu$A. The SQUIDs comprising the Faris pulsers are identical to the SQUID DUT. The strobe and DUT signals are triggered using fast current pulses carried to the sampler circuit on high-bandwidth lines (coaxial cables off-chip, CPW on-chip). Further fabrication details may be found in Appendix~\ref{app:fab}. Including all coupling resistors ($R_{sh,eq} = 0.9~\Omega$), and using fabrication parameters stated in Sec.~\ref{sec:sampler_sims}, we obtain an expected comparator $\beta_C = 4.56$ for the fabricated devices.

The sampler die is soldered using 52In/48Sn solder \cite{howe2012cryogen, schwall2011practical, chong2003thermal} inside a copper enclosure which is mounted on the second stage of a Gifford-McMahon-cooled cryostat and operated at 3.6~K. The device package possesses four high speed lines, used for applying flux trigger pulses, and 14 low-speed current bias and voltage tap leads. High speed lines exit the package via SMP connectors and are converted to SMA-connectorized semi-flex and rigid coaxial cables. We thermalize the flux trigger lines using 10~dB attenuators at 3.6~K. All low-speed lines are combined in a multi-wire cryogenic loom and are thermalized with two sets of 100~$\Omega$ series resistors at 3.6~K. Additionally a set of on-chip 100~$\Omega$ series resistors are fabricated as a final thermalization step and to create a stiff current bias in proximity to the sampler subcircuits. 

Our choice of Nb/\aSi/Nb JJs permits operation over a wide temperature range thanks to the temperature-insensitivity of $I_c$ for this JJ technology -- which is reduced to half its asymptotic low-temperature value (i.e. below 4~K) at around 7~K \cite{howe2022digital, olaya2023josephson}. Power dissipation ($P = I_cR_n$) becomes the primary consideration for sampler operation in terms of the low-temperature bound and limits operation of samplers using JJs with an $I_c$ of order 1~mA to around 1~K. Samplers using lower $I_c$ JJs may easily be constructed to permit operation at dilution refrigerator temperatures \cite{leonard2019digital, di2024control, yohannes2023high}.

\subsection{\label{sec:fab}Experimental Setup}
We apply current biases using a series of low-speed (1~GSa/s) AWGs configured to output square voltage pulses at a frequency of 20~kHz, with edge times of 30~$\mu$s, and a duty cycle of 0.8. The strobe and DUT flux triggers are supplied using step function voltage pulses from a multichannel, high-speed AWG (65~GSa/s, 25 GHz analog bandwidth). Discarding the bandwidth limit of cabling and interconnects, this means the trigger pulse rise time is of order 15~ps. Delay (advancement) of the firing time of the DUT relative to the strobe is accomplished by applying a negative (positive) dc level to the flux trigger by means of a dc voltage source, bias tee, and mechanical phase shifter (delay line) \cite{wolf1985josephson}. In our case we apply flux offsets to advance the DUT while the strobe trigger time is held constant. As a dc flux bias reduces the critical current of the SQUIDs in the DUTs and strobe, this technique is limited in the maximum strobe-DUT \textit{traversal} of approximately \mbox{30--100}~ps; depending on the proximity of $I_b^\text{sig}$ to $I_c^\text{sig}$ (less traversal is possible the higher the DUT bias). For timing calibration and system tuneup we operate the strobe and DUT triggers on independent channels, allowing us to program arbitrary strobe-DUT delays and determine optimal alignment parameters for the flux trigger waveform data. In this configuration we treat the flux trigger AWG sample clock as a master timing reference and perform a delay calibration of the mechanical delay line. Next, and to eliminate channel-to-channel jitter, we move to single output configuration where the strobe and DUT flux triggers are split with a power divider at the AWG output (with the bias tee and delay line following the divider for the DUT trigger line).

Rather than adopt the analog approach of historical JJ sampler demonstrations using a slow feedback loop, we implement a fully digital system for acquiring DUT waveforms using our sampler device. The comparator voltage is captured on a 125~MSa/s digitizer and we implement the same binary search algorithm used in our simulations \cite{vanzeghbroeck2023josephson}. The data from the full sampling period is post-processed to select the trace segment known to follow the arrival of the flux triggers, and use this voltage to determine if the comparator has switched for a given bias. Implementation of the binary search is thus straightforward with the single caveat that, in the presence of noise, we must collect multiple sampling events and enforce a switching probability threshold for each programmed strobe-signal delay (dc flux offset). Empirically, we find 1000 sampling events and a switching probability threshold of 0.5 to provide high-quality, repeatable sampled waveforms. With these settings and 10 binary search steps, the acquisition time is $\sim 1$~s per commanded delay -- limited by data transfer and post-processing. The advantage of this technique is that it directly mirrors the sampling process as performed in simulation, facilitating strong agreement between simulated and measured results.

\subsection{\label{sec:dut_Ib_sweep}Sampled Waveforms: DUT Current Bias Sweep}
\begin{figure}[tb]
    \centering
    \includegraphics[width = .48\textwidth]{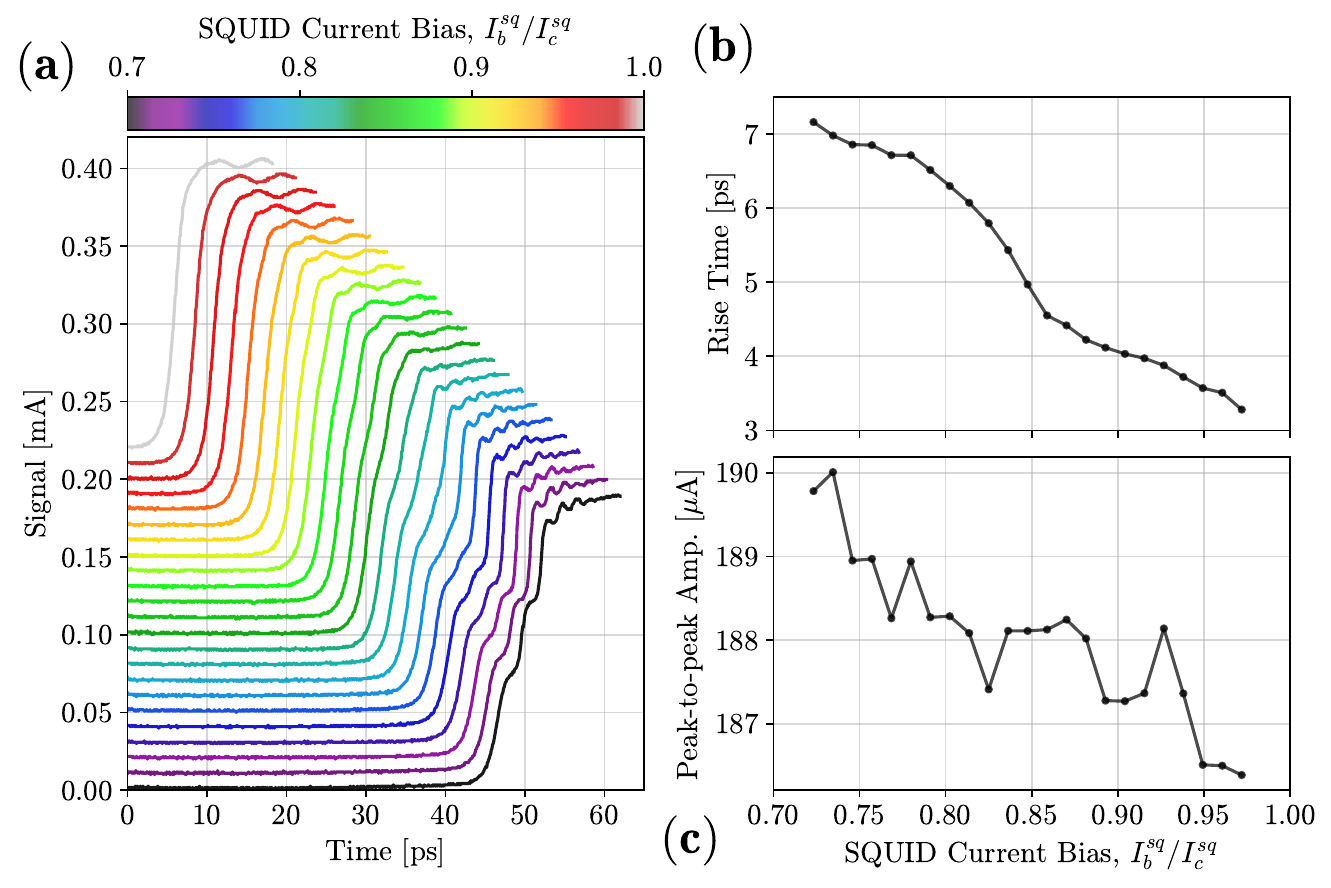}
    \caption{ Measured SQUID DUT signal and characteristics as a function of SQUID DUT bias, $I_b^{SQ}$. The strobe bias is fixed at 2.2~mA, slightly below its $I_c$ of 2.21~mA. \textbf{(a)} Sampled waveforms with each successive curve offset vertically by 10~$\mu$A for clarity. The time at which the SQUID latches moves to earlier times (smaller strobe-to-signal delay) as $I_b^{SQ}$ increases, so this horizontal shift is real. As the SQUID bias is ramped and crosses $I_b^{SQ} /I_c^{SQ} \approx 0.85$ we observe a transition between two distinct oscillation frequencies of $\sim 580$~GHz to $\sim 180$~GHz. \textbf{(b)} Measured sampler rise time as a function of the SQUID DUT bias current. \textbf{(c)} Sampled SQUID DUT signal amplitude as a function of the DUT bias current.}
    \label{fig:squid_Ib_sweep}
\end{figure}

\begin{figure}[tb]
    \centering
    \includegraphics[width = .48\textwidth]{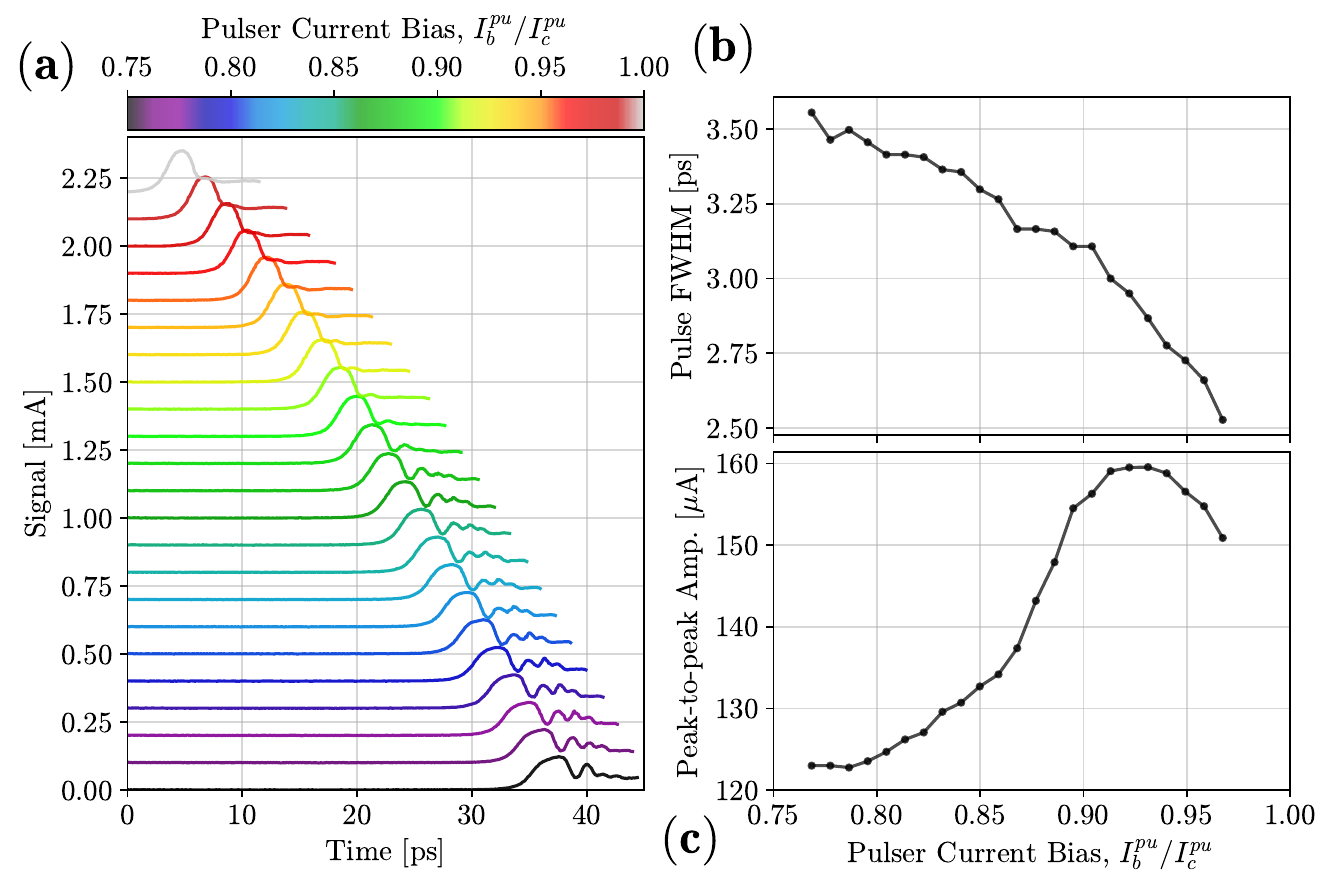}
    \caption{ Measured Faris pulser DUT signal and characteristics as a function of pulser DUT bias, $I_b^\text{pu}$, using a strobe bias of 2.2~mA. \textbf{(a)} Sampled waveforms with each successive curve offset vertically by 10~$\mu$A for clarity. As in Fig.~\ref{fig:squid_Ib_sweep} the horizontal shift of each curve is a physically real effect, i.e. the pulse is emitted at earlier times as $I_b^\text{pu}$ is increased.
    \textbf{(b)} Measured pulse FWHM for the sampled waveforms vs. normalized pulser DUT bias current $I_b^\text{pu} / I_c^\text{pu}$. \textbf{(c)} Sampled pulser DUT signal amplitude as a function of $I_b^\text{pu}$.}
    \label{fig:pulser_Ib_sweep}
\end{figure}

After initial tune-up and dc characterization of the device subcircuits (Appendix~\ref{app:dc_measurements}), we perform a series of sweeps to fully explore the sampler and DUT dynamics. We begin by selecting a fixed strobe current bias of 2.2~mA; close to its practical maximum of $I_c^\text{str} = 2.21$~mA to provide the sharpest strobe pulse \cite{donnelly20201ghz, vanzeghbroeck2024switching}. Noise- and thermal-oscillation-induced spurious switching is not present in our sampler even with the strobe bias this close to its critical current via sufficient filtering of the bias lines and the negligible temperature sensitivity of Nb/\aSi/Nb at 3.6~K \cite{olaya2023josephson}. We then sample the SQUID DUT waveform for a variety of DUT current biases. All of the sampler measurements presented in Fig.~\ref{fig:squid_Ib_sweep} and Fig.~\ref{fig:pulser_Ib_sweep} were collected by advancing the DUT relative to the strobe.

Calibration of the delay as a function of the flux trigger offset voltage (i.e. in ps/V corresponding to the signal advancement as depicted in Fig.~\ref{fig:sampler_ball_and_stick}(b)) is performed as in [\citen{wolf1985josephson}] using a mechanical delay line (phase shifter). This procedure involves sampling the DUT (at fixed current bias) waveform at two different delay line settings and measuring the separation between the curves, in flux offset voltage units, corresponding to the known delay induced by the two shifter positions. Using this technique the delay, per volt applied as a flux offset, may be known to sub-picosecond levels. This delay calibration is linear and reproducible with respect to the full range of applied DUT trigger dc flux offsets used to gather the sampled waveforms shown in Fig.~\ref{fig:squid_Ib_sweep} and Fig.~\ref{fig:pulser_Ib_sweep}. We obtain a constant value of 17.8 ps/V using a 1~k$\Omega$ current-defining resistor. See Fig.~\ref{fig:str_Ib_sweep_cal} in the appendix for more details.

In Fig.~\ref{fig:squid_Ib_sweep} and Fig.~\ref{fig:pulser_Ib_sweep}, the sampled waveforms each have approximately 600 points over a span of \mbox{40--60}~ps. At each point the sampler performs a binary search depth of 10, where each binary step consists of $10^3$ comparator decisions, for a total of $10^4$ comparator decisions made for each signal data point. The full scale search range selected for these signal data was 300~$\mu$A, which gives a current resolution of $\sim 300$~nA. To characterize repeatability and uncertainty of these measurements, we sampled and recorded several complete signal waveforms with identical settings for all control parameters. At each programmed delay we calculate the Type A uncertainty \cite{NIST1900} for the signal values. On the steepest regions of the curves corresponding to the largest uncertainty, we determine the DUT current Type A uncertainty for a 95\% confidence interval ($\pm 2\sigma$) of $\pm$1.2~$\mu$A ($k=2$). This is a relative uncertainty of $\sim0.6\%$ in Fig.~\ref{fig:squid_Ib_sweep} and $\sim1.2\%$ in Fig.~\ref{fig:pulser_Ib_sweep}. The error in the commanded delay arises from the uncertainty in the delay-to-dc-flux-offset calibration applied to the DUT flux trigger line. These uncertainties result in an expansion or compression of the waveforms by less than 0.3\%.

\begin{figure}[tb]
    \centering
    \includegraphics[width = .48\textwidth]{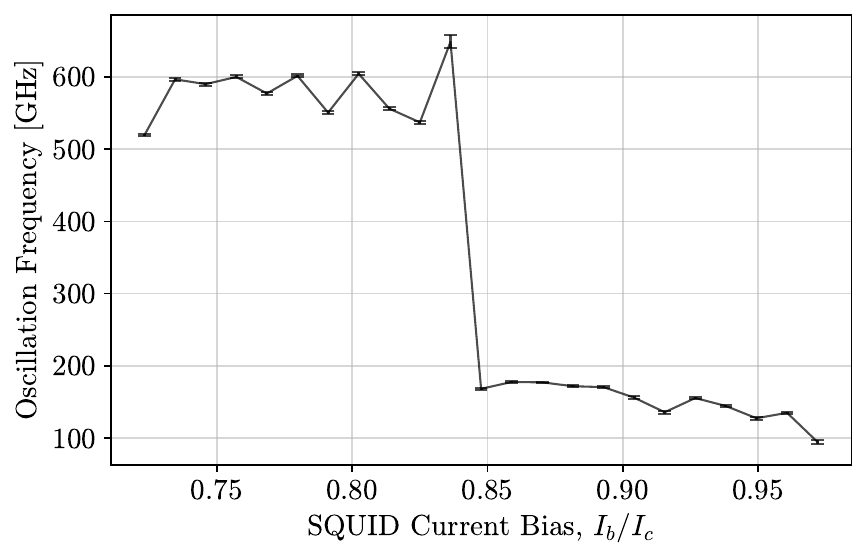}
    \caption{ Extracted oscillation frequency as a function of the SQUID DUT bias current. The frequency is determined via a fit to the signal above the 90\% level using Eq.~(\ref{eq:poly_sine}). Error bars correspond to the fit standard error for $f$.}
    \label{fig:squid_oscillation_freq}
\end{figure}

The family of sampled waveforms for the latching SQUID are shown in Fig~\ref{fig:squid_Ib_sweep}. At low DUT bias the sampled waveform shows two “bumps” in the rising edge of the step waveform that disappear with increasing $I_b^{SQ}$. Fast oscillations near the top of the rising edge also disappear and are replaced with slower oscillations. We may quantify the frequency of these oscillations by performing a polynomial and sinusoidal fit to the signal above its 90\% threshold via
\begin{equation}
    S(t) = c_0 + c_1 t + c_2 t^2 + c_3 t^3 + c_4 \sin(2 \pi f t + \phi).
    \label{eq:poly_sine}
\end{equation}
Fig.~\ref{fig:squid_oscillation_freq} shows the behavior of the SQUID DUT's oscillations as a function of its current bias. From this we see the comparator responds to oscillating signals with periods shorter than the extracted rise time of $\sim3.5$~ps; demonstrating our sampler is sensitive to signals in excess of 600~GHz. In this case, the measured value of the sampler $t_r$ is in fact partially limited by the SQUID DUT dynamics and its intrinsic rise time (as discussed in Sec.~\ref{sec:high_Jc_sim}).

These results, notably for DUT current biases of $I_b^{SQ} / I_c^{SQ} > 0.85$ are in good agreement with oscillations sampled in [\citen{wolf1985josephson}]. The physical dynamics giving rise to the two oscillation frequencies arise from the small delay in latching time of each of the SQUID junctions -- which in turn is a function of the SQUID bias and decreases as the bias increases (see Appendix~\ref{app:squid_latching_dynamics}). This time delay, ranging from 1.1~ps to 0.7~ps (from low to high bias), causes an appreciable beat oscillation between each SQUID JJ, which is detectable by our sampler even though this frequency is above the sampler bandwidth as extracted using a linear systems treatment. At high bias, the observed oscillations are reduced in frequency and are due to intrinsic oscillations of the underdamped comparator.

Next we perform the same measurement as above, but this time with the Faris pulser DUT. The strobe bias is again held at 2.2~mA and we sample the pulser DUT signal as a function of its bias current $I_b^\text{pu}$. These results, in tandem with the fast oscillations shown in Fig.~\ref{fig:squid_Ib_sweep}(a), demonstrate the intrinsic response time of the sampler is actually faster than what is extracted via the \mbox{10\%--90\%} rise time while sampling the SQUID DUT. Again we observe features on faster timescales than the minimum $t_r$ shown in Fig.~\ref{fig:squid_Ib_sweep}(b) and whose dynamics change as a function of the impulse DUT bias -- demonstrating they are sourced by the DUT and are not resulting from the convolution of the comparator dynamics. This is further evidenced by the fact the measured pulser DUT amplitude increases, reaches a measured maximum, and then begins to decrease -- while the FWHM monotonically decreases with increasing $I_b^\text{pu}$. Similar to the variation in the sampled oscillation for the step signal DUT, the reduction in FWHM is the result of a reduction in the time delay between the switching of the two JJs comprising the SQUID of the Faris pulser. As the time between each JJ switching event is minimized, the pulse amplitude is expected to monotonically increase. Our observation of a maximum pulse amplitude at $I_b^\text{pu}/I_c^\text{pu} = 0.925$, corresponding to a DUT FWHM of 2.8~ps, shows this is the regime where the intrinsic speed of the comparator is exceeded by the speed of the pulser DUT signal. Beyond this point the reduction in the sampled DUT FWHM is expected as, at this point, the sampler has entered the regime of a bandwidth-limited detector.

\begin{figure}[t]
    \centering
    \includegraphics[width = .48\textwidth]{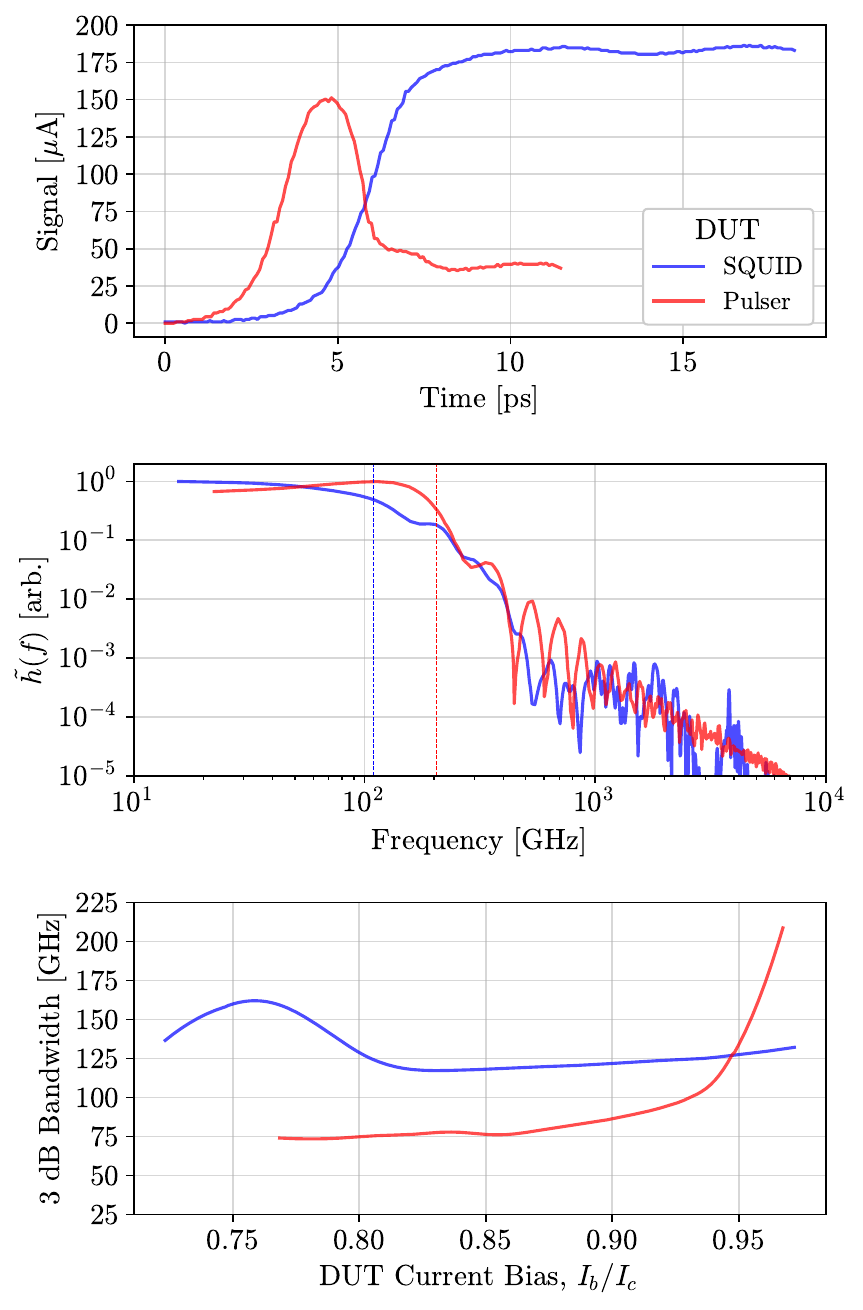}
    \caption{ \textbf{(a)} Highest DUT current bias sampled waveforms of Fig.~\ref{fig:squid_Ib_sweep} and Fig.~\ref{fig:pulser_Ib_sweep} corresponding to $I_b / I_c$ of 0.98 (0.97) for the SQUID (pulser). \textbf{(b)} Measured convolved sampler transfer function, $\tilde{h}(f)$, (in units of power spectral density scaled to the dc power) using the Fourier transform of the two traces in (a). Dashed lines denote the extracted 3~dB bandwidth in each case. This measurement yields an estimate of the sampler intrinsic transfer function as it provides the convolution of the DUT signal with the sampler transfer function. \textbf{(c)} Dependence of the 3~dB bandwidth on the DUT current bias.}
    \label{fig:transfer_func_and_3dB_bw}
\end{figure}

\section{\label{sec:bw_analysis}Sampler Bandwidth}
The 3~dB bandwidth, $B_{3~\mathrm{dB}}$, of the sampler may be approximated using the measurements in the previous section by treating the sampler as a first-order linear system \cite{aastrom2007feedback}. We perform this analysis on the data in Fig.~\ref{fig:squid_Ib_sweep} and Fig.~\ref{fig:pulser_Ib_sweep} where the strobe current bias is held fixed as each DUT current bias is swept. To extract $B_{3~\mathrm{dB}}$ we compute the transfer function corresponding to each $I_b^\text{str}$ and $I_b^{SQ}$ trace by Fourier transforming the sampled response and locating the point where the transfer function drops by 3~dB from its dc value. In our case, we choose the dc value to be the lowest frequency in the transfer function. Due to the steep rolloff in the transfer function no windowing is necessary to suppress ringing at high frequencies. As the SQUID DUT traces are of non-uniform length the data must be zero-padded so each record has a length equal to the longest un-padded trace. Additionally, because the transfer function applies only to periodic (or finite-duration) pulses, we differentiate the SQUID DUT data after zero-padding. Neither procedure is needed for the pulser DUT bandwidth extraction \cite{aastrom2007feedback}.

Fig.~\ref{fig:transfer_func_and_3dB_bw} shows the results of this analysis. Note that, because the DUT signals are not infinitely sharp, these measurements extract the convolution, $\tilde{h}(f)$, of the DUT signal with the sampler's true transfer function, $h(f)$. We observe that $B_{3~\mathrm{dB}}$ is clearly not saturating at the highest DUT biases (narrowest pulses and fastest rise time step signals), which demonstrates a lower bound of 175~GHz for the intrinsic sampler bandwidth. This agrees well with the fitted oscillation frequencies in the sampler reconstruction of the SQUID DUT waveforms as a function of DUT bias (Fig.~\ref{fig:squid_Ib_sweep}) and, were it possible to realize a sharper Faris pulser DUT pulse, would be expected to display a rolloff near 600~GHz. While it is true that the system dynamics are still increasing even at the highest DUT biases measured, it is not possible to further increase the bias and still fully sample the target waveform. This is because the dc flux offset delay technique also reduces the DUT critical current and, for a given $I_b / I_c$, this sets a limit on how much the DUT may be advanced before it latches during the set bias phase of the sampling period.

\section{Conclusions}
We have performed detailed simulations using a novel binary search threshold detection algorithm to understand the critical design criteria for optimizing the performance of a JJ-based sampler realized via a single latching comparator junction and two galvanically-connected DUTs: a latching SQUID and Faris pulser. The JJ parameters chosen in simulation closely match those observed using NIST's 0.2~\mAumsq~$J_c$ Nb/$\aSi$/Nb JJ process \cite{olaya2023josephson}. Using the latching SQUID DUT to provide a step signal we establish bounds on the expected intrinsic sampler rise time to be approximately \mbox{2.4--3.4~ps}. With the Faris pulser we find the minimum pulse FWHM the sampler can resolve is \mbox{2.5--3.0~ps}. Via full simulation of the sampling process we reveal a tradeoff between a small increase in the sampler speed at the expense of significant ($\gtrsim$10\% peak signal amplitude) distortion when pushing to the lower limit of $t_r$ (FWHM) by increasing the comparator $\beta_C$. Exploration of the relative coupling strengths of the strobe pulse and DUT signal, $r$, showed a relative narrow optimum range of $r \in [1, 4]$. In these simulations we note that the fastest response (minimum $t_r$ and FWHM) is, in some cases, limited by the DUT dynamics and not those of the comparator. Using a SQUID DUT model optimized to provide a 0.47~ps 10\%--90\% rise time, we demonstrate the comparator speed limit is actually expected to be 1.9~ps for the 0.2~\mAumsq~process. We also simulated capacitive and inductive DUT-comparator coupling schemes (Appendix~\ref{app:reactive_coupling}) to determine upper bounds on the coupler capacitance and inductance below which the sampler performance (bandwidth, signal distortion) is not significantly compromised.   

As an experimental realization of JJ sampler technology, we have designed and fabricated a latching Josephson sampler galvanically-connected to two DUTs using our modern JJ tri-layer process with niobium electrodes and amorphous silicon barriers (Nb/a-Si/Nb). Waveforms of both DUTs were experimentally measured at 3.6~K using the same binary search technique employed in simulation, each as a function of DUT and strobe bias. Our experimental results are in good agreement with simulation and showed the fastest response corresponding to a rise time and FWHM of 3.3~ps and 2.5~ps when measuring the SQUID and Faris pulser DUTs, respectively. These values correspond to the rise time and FWHM for sampled waveforms with the sampler operating in a distortion-minimizing state (high strobe bias, see Appendix~\ref{app:strobe_Ib_sweep}). A linear-systems analysis of the sampled SQUID and pulser waveforms yield a sampler bandwidth of 160~GHz and 210~GHz, respectively. However, we measure oscillatory signals above 600~GHz which transition to a lower frequency as DUT bias is increased. This dependence on DUT bias indicates the oscillations source from the DUT itself and are thus a real component of the signal rather than an artifact of the comparator dynamics convolved with the true signal. Observation of these high frequency signals both supports the assertion that our characterization of the intrinsic sampler speed limit is limited by the DUT circuits, not the comparator, and demonstrates potential bandwidth capabilities well beyond the previous best-in-field demonstrations. This simulation framework and laboratory demonstration provides a concrete path forward for ultra-fast and high-precision on-chip waveform metrology for superconducting electronics.

\begin{acknowledgments}
L. Howe thanks M. Castellanos-Beltran for numerous fruitful discussions lending greatly to the success of this work. This work is a contribution of the U.S. government and is not subject to U.S. copyright. 
\end{acknowledgments}

\section*{Data Availability Statement}
The data that support the findings of this study are openly available in \textit{Josephson Samplers: Optimal Design and Demonstration} at \url{https://datapub.nist.gov/od/id/mds2-3649}, with DOI doi:10.18434/mds2-3649.

\bibliography{references}% Produces the bibliography via BibTeX.

\begin{thebibliography}{56}
\bibitem [{NIS()}]{NISTdisclaimer}%
  \BibitemOpen
  \href@noop {} {}\bibinfo {note} {Commercial instruments and software are
  identified in this paper in order to adequately specify the experimental
  procedure. Such identification does not imply recommendation or endorsement
  by {NIST}, nor does it imply that the product identified is necessarily the
  best available for the purpose}\BibitemShut {NoStop}%
\end{thebibliography}

%%%%%%%%%%%%%%%%%%%%%%%%%%%%%%%%%%%%%%%%%%%%%%%%%%%%%%%%%%%%%%%%%%%%%%%%%%%%%%%%%%%%%%%%%
%%%%%%%%%%%%%%%%%%%%%%%%%%%%%%%%%%%%%%%%%%%%%%%%%%%%%%%%%%%%%%%%%%%%%%%%%%%%%%%%%%%%%%%%%
%%%%%%%%%%%%%%%%%%%%%%%%%%%%%%%%%%%%%%%%%%%%%%%%%%%%%%%%%%%%%%%%%%%%%%%%%%%%%%%%%%%%%%%%%
%%%%%%%%%%%%%%%%%%%%%%%%%%%%%%%%%%%%%%%%%%%%%%%%%%%%%%%%%%%%%%%%%%%%%%%%%%%%%%%%%%%%%%%%%
%%%%%%%%%%%%%%%%%%%%%%%%%%%%%%%%%%%%%%%%%%%%%%%%%%%%%%%%%%%%%%%%%%%%%%%%%%%%%%%%%%%%%%%%%

\clearpage
\appendix

% Second title page
\begin{titlepage}
  \centering
  \vskip 60pt
  \LARGE Picosecond Josephson Samplers: Modeling and Measurements -- Supplemental Information \par
  \vspace{5mm}
\end{titlepage}

\section{\label{app:binary_search}Binary Search Algorithm and Implementation}
The binary search algorithm \cite{vanzeghbroeck2023josephson} starts with the selection of a minimum and maximum bias current that will be applied to the comparator. These are to be chosen such that the range of amplitudes in the signal of interest is smaller than the difference between those two currents and with an appropriate offset to avoid clipping of the response. Properly chosen, the max (min) biases will result in the comparator always (never) latching for all delays -- i.e for all delays which align the strobe to sample any portion of the signal.

First, two initial simulations are run with $I_b^\text{comp} = I_\text{max}^\text{comp}$ and $I_b^\text{comp} = I_\text{min}^\text{comp}$ to guarantee the maximum and minimum biases are appropriately chosen. We then define, for iteration number $n$,
\begin{equation}
    I_\text{high}^{n=0} =  I_\text{max}
\end{equation}
\begin{equation}
    I_\text{low}^{n=0} = I_\text{min},
\end{equation}
\begin{equation}
    \Delta I_b^{n+1} = \frac{I_\text{high}^n - I_\text{low}^n}{2},
\end{equation}
set the bias for the first binary search step to
\begin{equation}
    I_b^{n=1} = I_\text{min} + \Delta I_b^{n=1} = I_\text{min} + \frac{I_\text{max} - I_\text{min}}{2},
\end{equation}
and determine if the junction latches under this dc bias (a ``1") or not (a ``0"). If the result of simulation $n$ was a 1 we select the lower bias midpoint by setting
\begin{equation}
    I_\text{high}^{n+1} = I_b^n
\end{equation}
\begin{equation}
    I_\text{low}^{n+1} = I_\text{low}^n,
\end{equation}
and conversely for simulation $n$ yielding the 0 result we select the upper midpoint:
\begin{equation}
    I_\text{high}^{n+1} = I_\text{high}^n
\end{equation}
\begin{equation}
    I_\text{low}^{n+1} = I_b^n.
\end{equation}
Thus, the bias at simulation step $n\geq1$ is
\begin{equation}
    I_b^{n+1} = I_b^{n} + \Delta I_b^{n+1} = I_b^n + \frac{I_\text{high}^n - I_\text{low}^n}{2}.
\end{equation}

We implement this technique in simulation using WRspice \cite{NISTdisclaimer,wrspice}. To simulate a full sampling reconstruction of the target waveform with respect to a given circuit parameter, we simulate only single sampling periods at a time, of duration \mbox{200--1200~ps} depending on the parameter sweep being run, and adjust only the comparator bias or signal trigger delay. We first fix the signal trigger delay and perform a binary search to determine the new value at which $I_b$ results in latching of the comparator. Latching of the comparator is determined when the mean voltage over a fixed time window exceeds a threshold (typically $\sim 1/2$ the gap voltage). The number of steps in the binary search is chosen to yield an $I_b$ resolution of $\leq 200$~nA. Next the signal trigger delay is adjusted, we repeat the $I_b$ search, and iterate until the entire target waveform is sampled. Finally, to characterize the response of the full sampler output with respect to a given circuit parameter, we adjust this parameter and repeat the full sampling operation described above.

Simulations of the sampling process using the binary search are significantly faster than simulations that implement the slow (of order \mbox{10--100~Hz}) feedback circuit method used in the experiments by \citen{wolf1985josephson}. The speedup can be as large as a factor of $10^4$ or more. Furthermore, our binary search simulation strategy yields high-resolution ($\delta I \sim 100$~nA, $\delta t \sim 0.1$~ps) simulations of complete target waveforms in minutes \cite{vanzeghbroeck2023josephson}.

\begin{figure}[t]
    \centering
    \includegraphics[width = .48\textwidth]{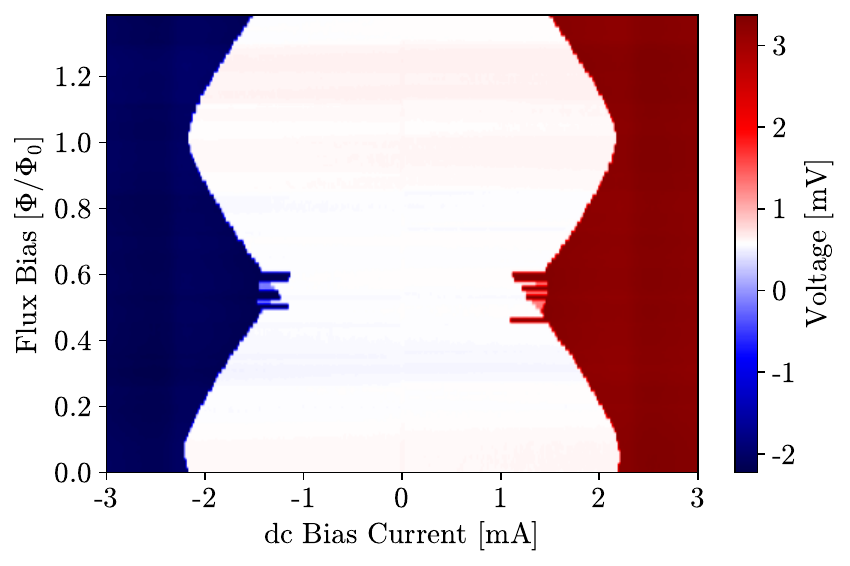}
    \caption{ Critical current of the Faris pulser strobe SQUID as a function of the dc flux bias applied to the flux trigger line.}
    \label{fig:Ic_vs_phi}
\end{figure}

\section{\label{app:dc_measurements} DC Subcircuit Characterization}

As shown in in the circuit diagram of Fig.~\ref{fig:sampler_pic_and_fabricated_device_schematic}(c), the designed sampler has current and voltage taps for each subcircuit on the sampler die (comparator, strobe, SQUID DUT, and pulser DUT). This feature allows us to easily characterize each individual component of the sampler via \mbox{$I$-$V$} and $I_c$ vs. $\Phi$ curves. One such curve for the strobe generator is shown in Fig.~\ref{fig:Ic_vs_phi} where a dc flux is applied via the strobe's high speed flux trigger line, and the bias current is then ramped up and down from zero to measure $I_c^\text{str}$. We repeat these measurements for the comparator (only an \mbox{$I$-$V$} curve as there is no flux control of the comparator), SQUID, and Faris pulser DUTs to determine appropriate bias currents and flux pulse amplitudes. Note that the effect of applying a flux pulse to trigger waveform emission is to traverse the critical current in the \textit{vertical} direction in Fig.~\ref{fig:Ic_vs_phi}.

As an additional characterization we confirm the design of the SQUIDs internal to all on-chip signal generators by measuring the screening parameter and extracting the SQUID loop inductance. The SQUID screening parameter $\beta_L = 2 L_{SQ} I_c / \Phi_0$ is estimated \cite{vanzeghbroeck2023josephson} by ramping the dc flux bias through the flux trigger line while measuring the SQUID critical current, and using
\begin{equation}
    \frac{I_{c}^{min}}{2I_{c}^{max}} = \frac{\frac{\pi \beta_L}{2} + (\pi \beta_L)^2}{2 + \pi^2\beta_L + (\pi \beta_L)^2}.
\end{equation}
We obtain $\beta_L = 3.7$ and a total SQUID loop inductance of $L_{SQ} = 1.1$~pH. Electromagnetic simulations estimated this value to be $L_{SQ} = 1.6$~pH and we attribute this discrepancy to the apodization near the SQUID loop pinhole during photolithography -- resulting in a smaller open area than was designed. The designed geometry was a $1 \times 1~\mu$m square, which is near the minimum feature size for optical lithography in our fabrication process. From the micrograph in Fig.~\ref{fig:sampler_pic_and_fabricated_device_schematic}(a) we obtain a geometry closer to $0.8~\mu$m~$\times~0.8~\mu$m which can explain a lower screening parameter as a smaller effective SQUID loop lowers $\beta_L$.

%Our system displays nominal inductive crosstalk in the dc bias leads due to their close proximity inside the twisted pair cryolooms and we observe non-negligible current bias spikes in neighboring devices during the simultaneous bias current ramp up (down) at the beginning (end) of a sampling period. To circumvent this issue we capture the comparator voltage over an entire sampling period, which includes the crosstalked voltage due to the bias ramp, and simply post-process the data to select

\section{\label{app:fab} Device Fabrication}
\begin{figure}[tb]
    \centering
    \includegraphics[width=0.49\textwidth]{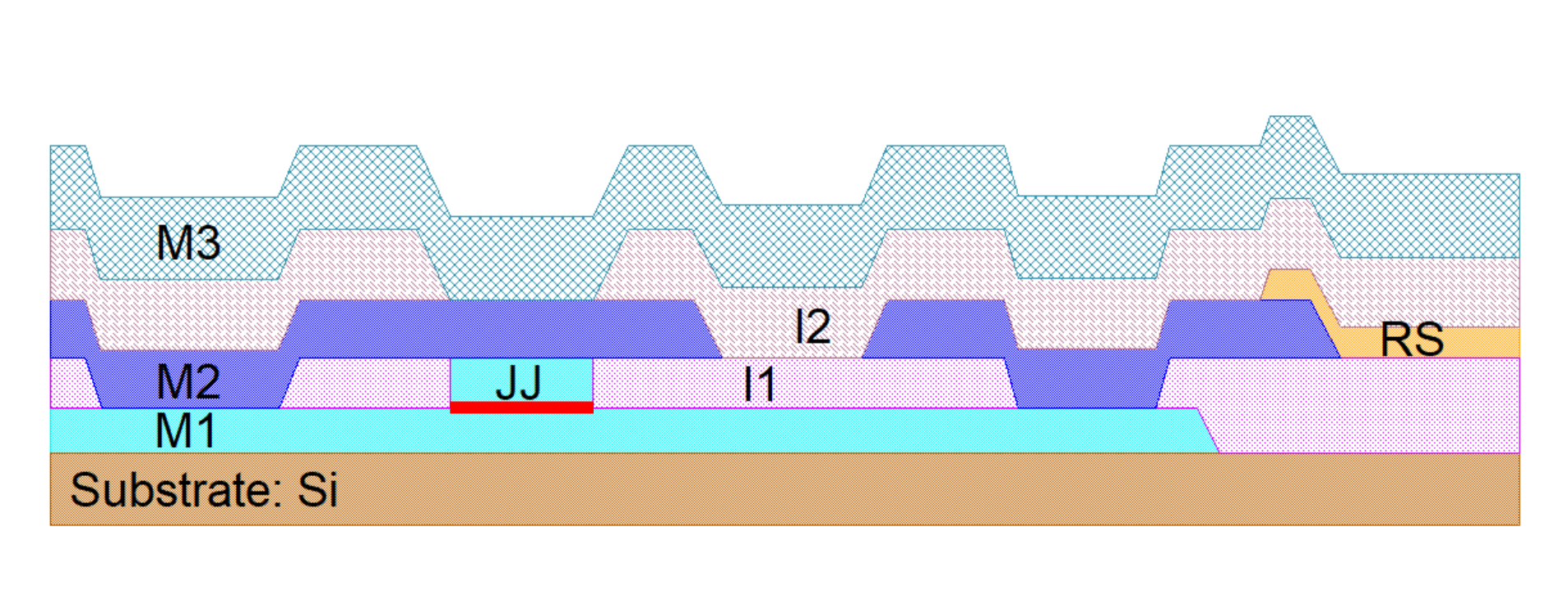}
    \caption{Schematic cross-sectional view of the JJ sampler layer stack. M\#: Nb layers, I\#: insulator layers, RS: resistor layer, JJ: Josephson junction. The barrier is denoted by the red line. The top metal layer (M3) is the sky ground plane.}
    \label{fig:fab_schematic}
\end{figure}

The fabrication process for the sampler circuits, shown in Fig.~\ref{fig:fab_schematic}, is based on the process for SFQ circuits developed at NIST and described in [\citen{olaya2019planarized}], with a few modifications. In this case, the substrate was 76~mm intrinsic silicon with no thermal oxide. Three superconducting Nb layers were used, M1 to M3 from bottom to top, all deposited by dc sputtering and patterned by reactive ion etching using sulfur hexafluoride gas. The junction barrier was sputtered amorphous silicon, deposited in situ with the base electrode (M1) and junction counter electrode (M2). The barrier thickness was chosen to obtain a junction critical current density of 0.2~\mAumsq. The insulating layers, consisting of silicon oxide, were deposited by electron cyclotron resonance plasma enhanced chemical vapor deposition (ECR-PECVD). Only one planarization step, on insulator layer I1, was done. The resistors, of palladium gold alloy and defined by lift-off, were deposited on top of M2 layer, as shown in the schematic.

\begin{figure*}[t!]
    \centering
    \includegraphics[width = .95\textwidth]{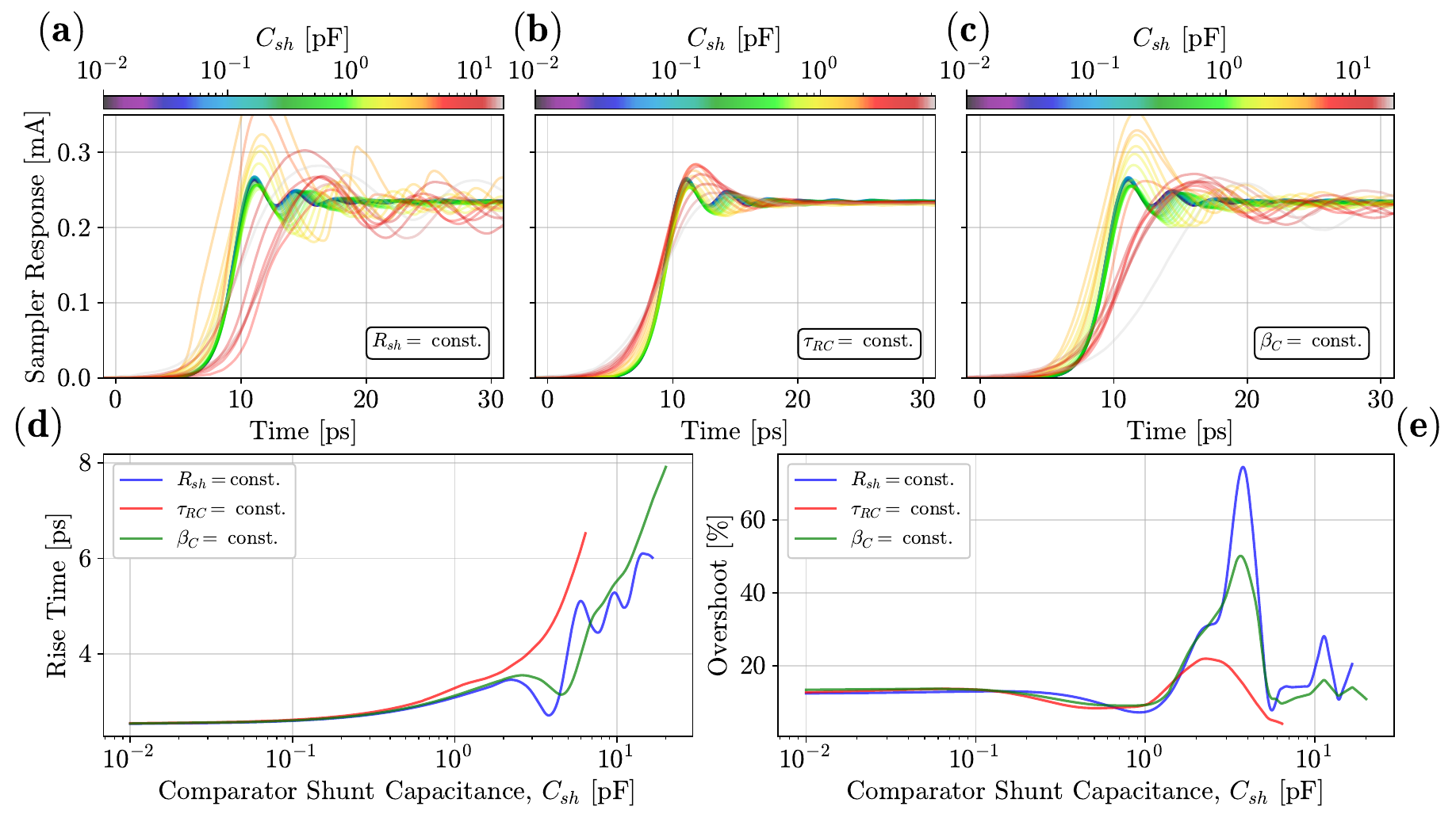}
    \caption{ Simulations of the sampler response with a capacitive DUT-sampler coupling as the coupling capacitor $C_{sh}$ is varied and with a baseline $R_{sh} = 5~\Omega$. We have the freedom to either tune $R_{sh}$ to maintain either a fixed comparator $\tau_{RC}$ or fixed $\beta_C$. Note that the linear dependence on the $RC$ time (as opposed to quadratic for $\beta_C$) restricts the range of $C_{sh}$ for the constant $RC$ time case relative to the other two cases. \textbf{(a)--(c)} Show the sampled SQUID DUT step signal waveforms when $R_{sh}$, $\tau_{RC}$, and $\beta_C$ are held constant, respectively. \textbf{(d)} and \textbf{(e)} respectively show the rise time and overshoot for the three cases.}
    \label{fig:Csh_t_res_sweep}
\end{figure*}

\section{\label{app:reactive_coupling}Reactive Coupling Simulations -- Detailed Results}
In some instances a galvanic comparator-DUT coupling may be undesirable, e.g. in the event a scanning sampler is the preferred system \cite{cui2017scanning}. For the topology of Fig.~\ref{fig:sim_schematic}, reactive coupling may be achieved either by a capacitance in parallel (shunt), $C_{sh}$, with the comparator junction, or with an inductance in series, $L_s$. First we examine the case of a parallel capacitive coupling by performing a $t_r$ simulation while sweeping $C_{sh}$. To keep the DUT output signal constant during the simulation, the galvanic connection was kept ($R_c^{\text{sig}} = 12~\Omega$) in order to enforce a fixed strobe-to-signal ratio and avoid any complications detailed in the results of Fig.~\ref{fig:tres_vs_Rcouple}. Note that in a true capacitively-coupled sampler the circuit topology would be the same as in Fig.~\ref{fig:sim_schematic} except that $R_c^{\text{sig}}$ would be replaced by a coupling capacitor $C_c^{\text{sig}}$; in this case the equivalent shunt capacitance of the comparator is $C_{sh}^{\text{eq}} = C_{sh} + C_c^{\text{sig}}$ (in this simplified simulation we ignored the capacitance to ground in the DUT). For typical coupling capacitances of $\sim 1$~pF and voltage slew rates of $\sim 1$~mV/ps, the peak current through the capacitor is $I^\text{sig} = C_c^{\text{sig}}dV/dT \sim 1$~mA -- demonstrating signals with sufficient amplitudes may be applied to the comparator via capacitive couplers.

The coupling capacitance is expected to have negligible effect on the sampler operation in the regime $C_{sh} \ll C_{JJ}^\text{comp} = 1$~pF. In these simulations, however, once $C_{sh}$ approaches the comparator junction capacitance it perturbs both the comparator Stewart-McCumber parameter and the $RC$ time constant $\tau_{RC}$. As we anticipate requiring couplers with close to 1~pF of capacitance we explore two options to compensate for these effects: \textit{(i)} adjust the explicit shunt resistance to keep $\tau_{RC}$ constant, and \textit{(ii)} adjust $R_{sh}$ to keep $\beta_C$ constant -- as well as the case where no compensation is performed. In both cases $R_{sh}$ must be reduced as $C_{sh}$ increases so we begin these simulations with a negligible $C_{sh} = 0.01$~pF and $R_{sh} = 5~\Omega$ (baseline $\beta_C = 9.9$) to permit compensation over the widest range of $C_{sh}$ while maintaining an underdamped comparator. Fig.~\ref{fig:Csh_t_res_sweep} shows the results of these simulations and, as expected, little impact is seen for $C_{sh} \lesssim 1$~pF. For $C_{sh}$ above 1~pF both the uncompensated and constant-$\beta_C$ case show adverse distortion relative to the constant-$\tau_{RC}$ case but do yield small gains in preserving the low-$C_{sh}$ rise time compared to the constant-$\tau_{RC}$ results. Nevertheless, all three parameter sweeps clearly demonstrate that, for optimal performance in a capacitively-coupled device, care must be taken to ensure the comparator junction capacitance is dominant.

\begin{figure*}[t!]
    \centering
    \includegraphics[width = .95\textwidth]{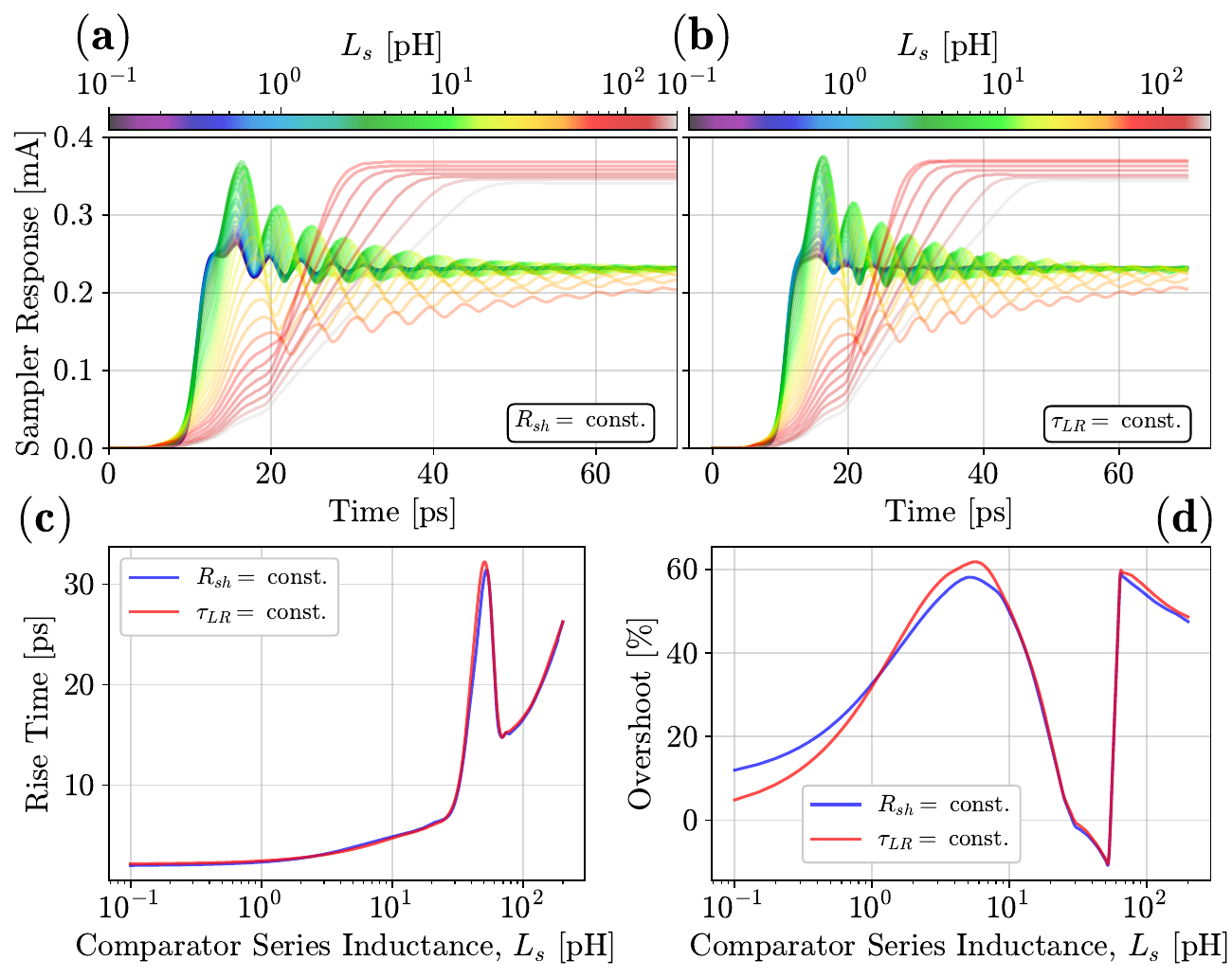}
    \caption{ Inductively-coupled sampler simulation results. \textbf{(a)} Constant $R_{sh} = 5~\Omega$ case. \textbf{(b)} Constant $\tau_{LR}$ case where $R_{sh}$ is initially set to 1~$\Omega$ and increased in tandem with $L_s$. \textbf{(c)} Rise time as a function of $L_s$. \textbf{(d)} Overshoot as a function of $L_s$.}
    \label{fig:Ls_t_res_sweep}
\end{figure*}

Next we consider inductive coupling via a series inductance (i.e. a pickup coil). As before, values of $L_s$ far below the comparator Josephson inductance $L_J^\text{comp} = \Phi_0 / 2 \pi I_c^\text{comp} = 0.16$~pH should not affect operation. Similar to the capacitive coupling case, we may choose to adjust $R_{sh}$ to compensate for shifts in the comparator $L/R$ time constant. In this case the greatest tunable range is yielded by starting with the \textit{lowest} $R_{sh}$ value that still enforces an underdamped comparator: $R_{sh} = 1~\Omega$. Results are shown in Fig.~\ref{fig:Ls_t_res_sweep} for both a constant shunt of $R_{sh} = 5~\Omega$ and for an adjusted shunt which maintains the $L/R$ time of the $R_{sh} = 1~\Omega$ case. The sampler performance is quite similar for both cases and, as in the capacitive-coupling case above, the onset of distortion and degradation occurs once the reactive element (coupling inductor) is no longer a small perturbation compared to the comparator. I.e., the rise time and overshoot are nearly constant until the coupling inductance approaches the comparator's Josephson inductance $L_J = \Phi_0 / 2 \pi I_c \sim 1.6$~pH. Increasing $L_s$ to approximately ten times the comparator $L_J$ results in a moderate reduction in rise time (with little to no impact in overshoot) but for $L_s \gtrsim 10 L_J$ the sampler's performance is severely impacted.

\section{\label{app:strobe_Ib_sweep}Sampled Waveforms: Strobe Current Bias Sweep}
\begin{figure}[t]
    \centering
    \includegraphics[width = .48\textwidth]{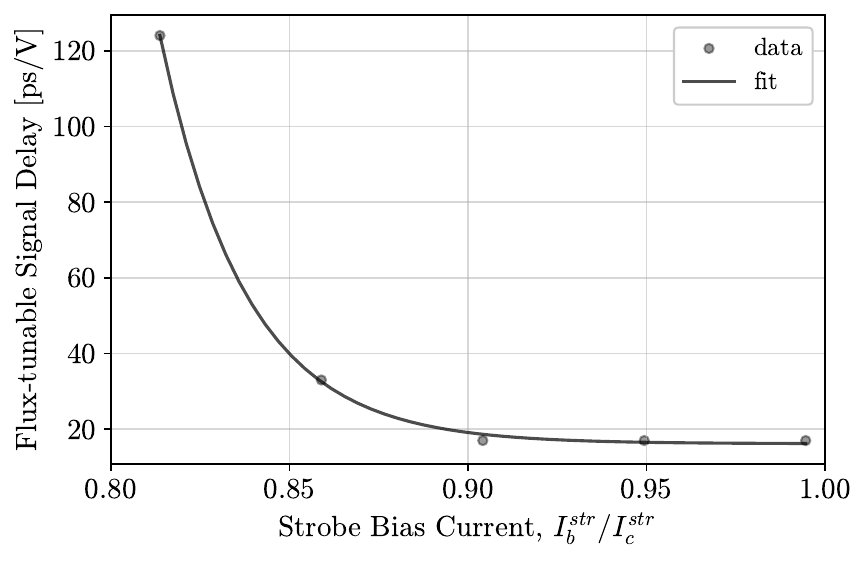}
    \caption{Calibration of the dependence of the signal advancement (per volt of dc flux offset applied at room temperature) as a function of the strobe bias $I_b^\text{str}$. Each data point is obtained by acquiring two DUT waveforms at different mechanical delay positions and fixed $I_b^\text{str}$ -- which provides a known and calibrated delay/advancement (in picoseconds) between the two curves. The separation between the two traces (in units of volts of applied flux offset) is then used to determine the calibration of the flux-tunable delay (units of picoseconds of advancement per volt of applied flux offset) at a given strobe bias. We observe an exponential trend as $I_b^\text{str}$ is adjusted and use a fit of this form for Figs.~\ref{fig:strobe_Ib_sweep_sample_squid} and \ref{fig:strobe_Ib_sweep_sample_pulser}.}
    \label{fig:str_Ib_sweep_cal}
\end{figure}

To fully characterize the sampler experimentally, we explore the sampler behavior not only by tuning the DUT current bias as in Sec.~\ref{sec:dut_Ib_sweep}, but also as we vary the strobe current bias, $I_b^\text{str}$ while holding the DUT biases fixed. I.e. the reverse scenario as covered in Sec.~\ref{sec:dut_Ib_sweep}. Strobe and signal alignment in this scenario is more challenging over an instructive range of $I_b^\text{str}$ due to a significant flux-offset-delay calibration nonlinearity. Indeed, the amount of DUT signal advancement (negative delay) per milliamp of dc flux bias applied to the DUT trigger line can be a strong function of the dc strobe current bias. Fig.~\ref{fig:str_Ib_sweep_cal} shows the results of calibrating this effect, performed at each strobe bias point using a tunable mechanical delay line and an exponential fit.

\begin{figure}[t]
    \centering
    \includegraphics[width = .48\textwidth]{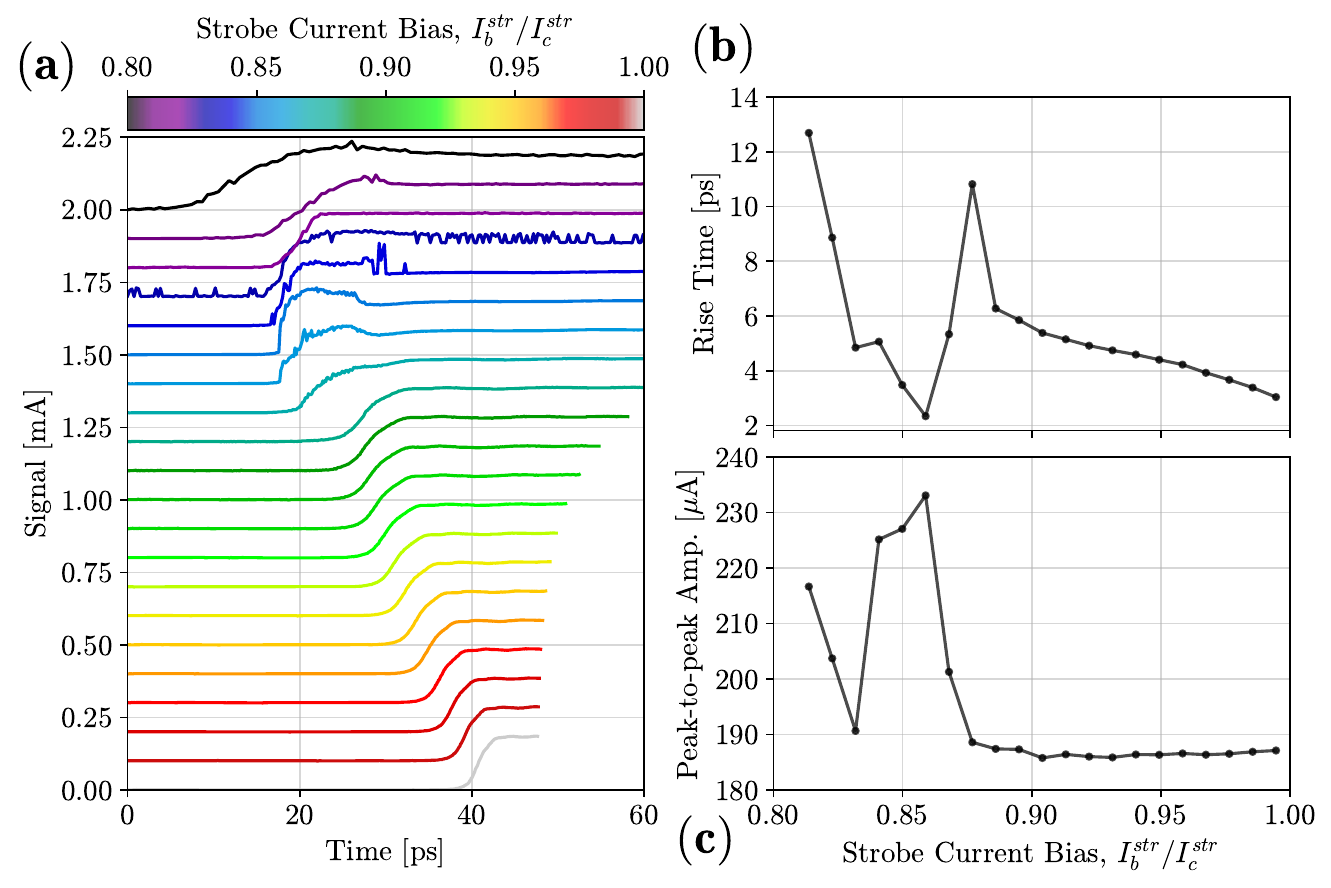}
    \caption{ Measured SQUID DUT response and characteristics as a function of strobe bias, $I_b^\text{str}$, using a SQUID bias of 2~mA ($I_b^{SQ}/I_c^{SQ} = 0.90$). \textbf{(a)} Sampled waveforms with each successive curve offset vertically by 10~$\mu$A for clarity. For low strobe bias the sampler fails to accurately reconstruct the target waveform and the result is a much slower response than expected with large distortion. Once $I_b^\text{str} / I_c^\text{str}$ exceeds $\approx 0.9$ the output is much closer to the expected signal from the SQUID DUT -- both in terms of the rise time and overall shape. \textbf{(b)} Extracted rise time vs. normalized strobe bias $I_b^\text{str} / I_c^\text{str}$. \textbf{(c)} Sampled SQUID DUT signal amplitude as a function of $I_b^\text{str}$.}
    \label{fig:strobe_Ib_sweep_sample_squid}
\end{figure}

\begin{figure}[h!]
    \centering
    \includegraphics[width = .48\textwidth]{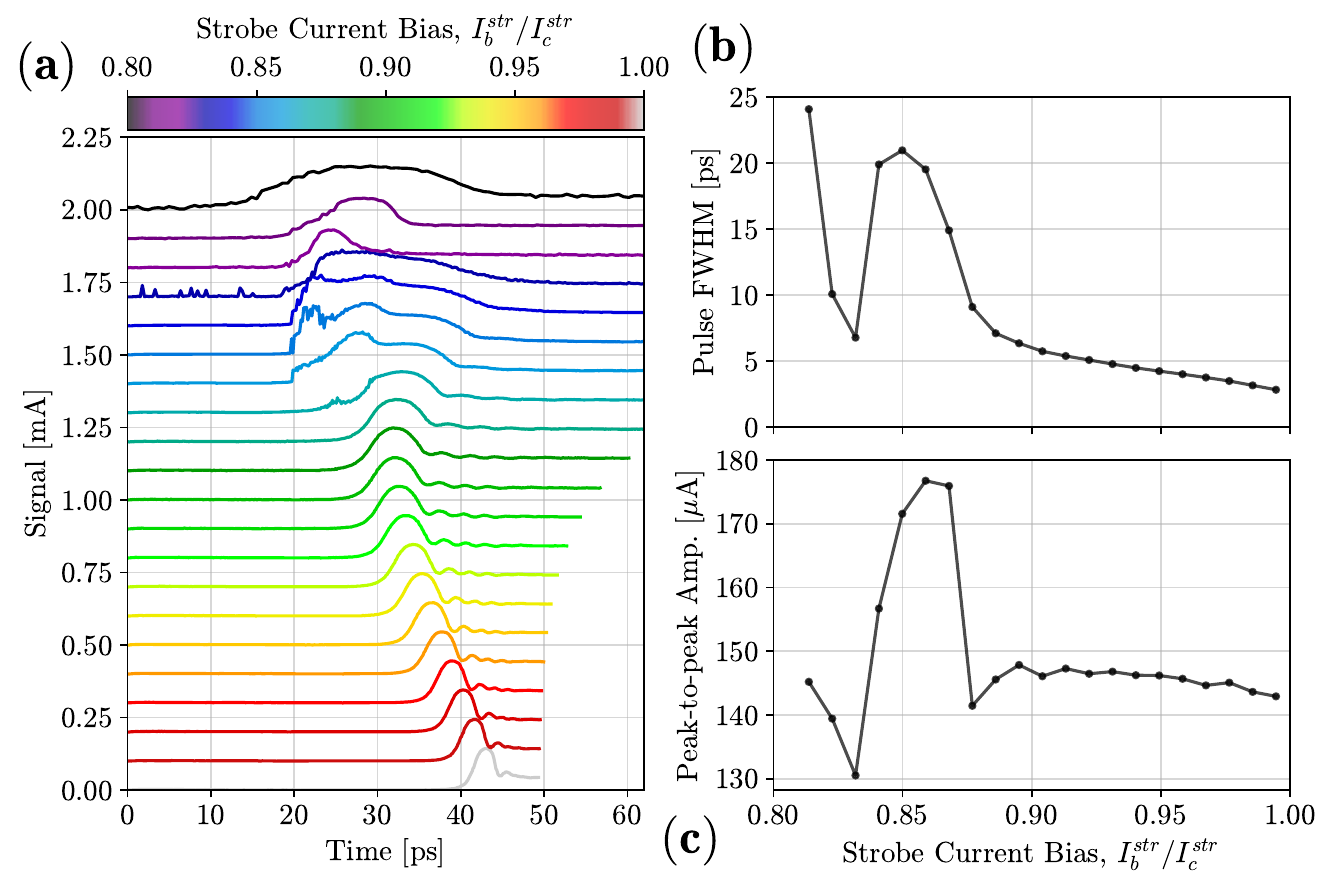}
    \caption{ Measured pulser DUT response and characteristics as a function of strobe bias, $I_b^\text{str}$, using a pulser bias of 1.95~mA ($I_b^{SQ}/I_c^{SQ} = 0.88$). \textbf{(a)} Sampled waveforms with each successive curve offset vertically by 10~$\mu$A for clarity. Again we see a clear transition -- as $I_b^\text{str}$ is increased -- from a heavily distorted and slow sampled signal, to a gradually sharper and low-distortion response as $I_b^\text{str}$ approaches $I_c^\text{str}.$ \textbf{(b)} Sampled pulse FWHM vs. normalized strobe bias $I_b^\text{str} / I_c^\text{str}$. \textbf{(c)} Sampled signal amplitude as a function of $I_b^\text{str}$.}
    \label{fig:strobe_Ib_sweep_sample_pulser}
\end{figure}

Fig.~\ref{fig:strobe_Ib_sweep_sample_squid} shows the results of sampling the SQUID DUT, with a fixed SQUID bias, while sweeping the strobe bias. Note the curve in Fig.~\ref{fig:strobe_Ib_sweep_sample_squid}(a) corresponding to $I_b^\text{str} / I_c^\text{str} = 0.86$ (starting y-axis value of 1.5~mA) and its resemblance to measurements in [\citen{wolf1985josephson}]. This also corresponds to both the shortest extracted rise time in our device of 3.3~ps and the maximal signal amplitude. However, this waveform displays appreciable distortion relative to the expected dynamics of an underdamped latching SQUID. The analogous measurement of the pulser DUT is shown in Fig.~\ref{fig:strobe_Ib_sweep_sample_pulser}. In both cases we observe that for the sampler to accurately reconstruct the target waveform the strobe bias must be $I_b^\text{str} / I_c^\text{str} \gtrsim 0.9$ and biases below this result in highly distorted sampled waveforms with slow rise (and fall) times. 

\begin{figure*}
    \centering
    \includegraphics[width = .95\textwidth]{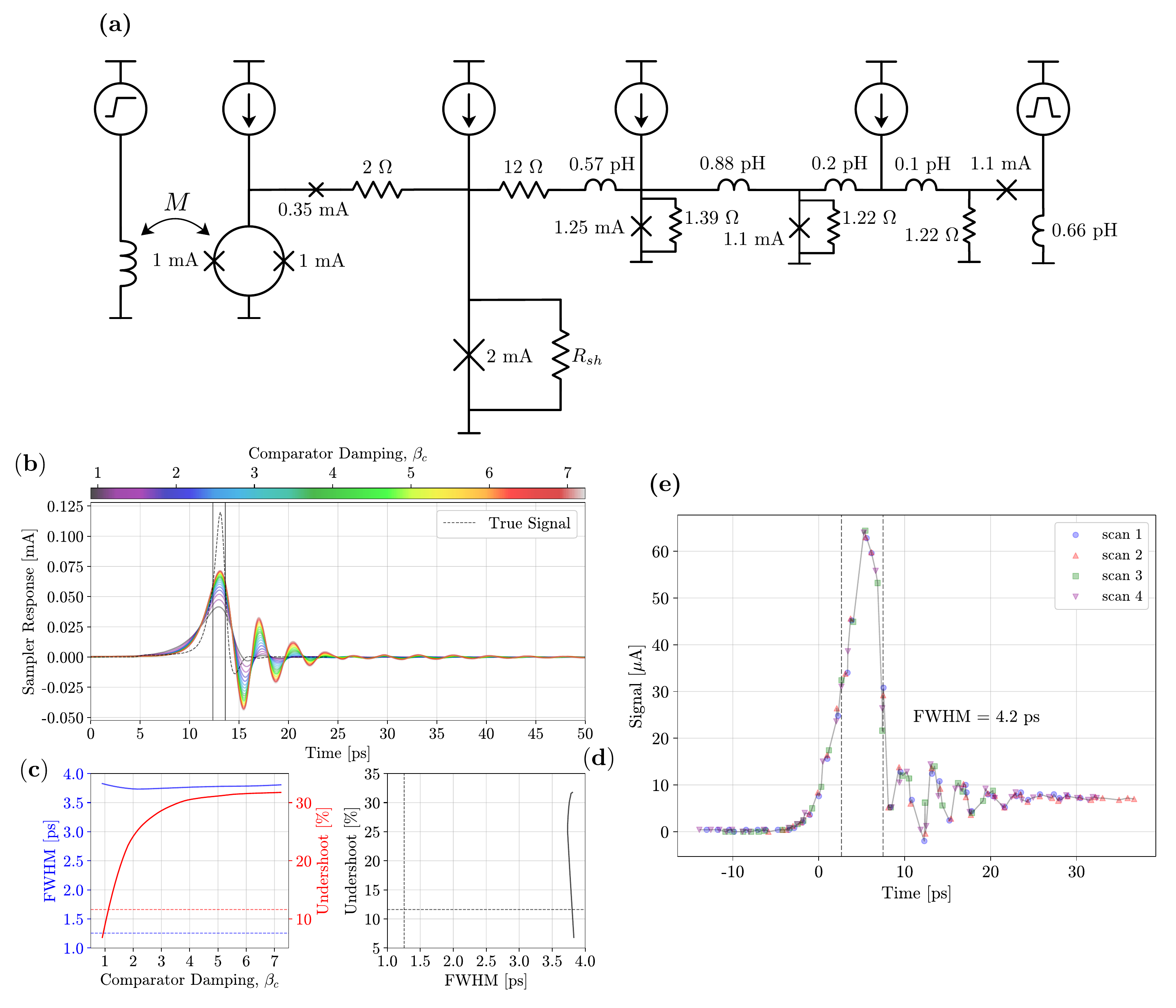}
    \caption{ Simulation and measurement of the sampled output of a DC/SFQ converter DUT. \textbf{(a)} Electrical simulation schematic.  The DC/SFQ converter is comprised of the elements to the right of the 12~$\Omega$ coupling resistor. All JJ specific capacitances are kept at the expected 80~fF/$\mu$m$^2$ as in the main text. For the DC/SFQ portion of the circuit we enforce critical damping of these JJs via addition of 1.39~$\Omega$ and 1.22~$\Omega$ shunts to the 1.25~mA and 1.1~mA JJs. \textbf{(b)} Simulated sampled SFQ pulse as a function of $\beta_C$ (solid colored lines). The dashed black line shows the simulated SFQ pulse at its output node while the vertical black lines show the half-maxima values of the simulated SFQ pulse. \textbf{(c)} Sampled pulse FWHM (blue) and undershoot (red). Dashed lines indicate the values intrinsic to the SFQ pulse. \textbf{(d)} Relation of the sampled pulse FWHM and its undershoot. The dashed vertical (horizontal) lines show the intrinsic FWHM (undershoot) of the SFQ pulse. Clearly evident in the simulation data shown in \textbf{(b)--(d)} is the expected behavior of a bandwidth-limited detector (i.e. the comparator) which measures a constant value of the SFQ pulse FWHM but as the detector bandwidth is lowered, the peak amplitude of the measured signal decreases due to its intrinsically-limited operating speed. \textbf{(e)} Sampler measurement of the DC/SFQ DUT at 3.6~K using a motor-controlled mechanical delay line to set the strobe-signal delay. We performed three repeated scans using the smallest achievable motor steps and removed spurious feedthrough of the trigger signal (see text). Lines connecting data points of the same color are guides for the eye.}
    \label{fig:sfq_pulse_vs_Rshunt}
\end{figure*}

\section{\label{app:sampling_dcsfq}Sampling SFQ Pulses}
As an extension to the simulation results detailed in Sec.~\ref{sec:faris_pulser_sim} we also simulate sampling of an alternate finite time-duration pulse DUT in the form of a DC/SFQ converter. Simulating and experimentally verifying that a Josephson sampler can measure the SFQ output of this type of circuit without significant distortion is most relevant to efforts in superconducting digital logic, computation, and signal generation performed using critically damped JJs. In designing the DC/SFQ converter we chose to maximize the amplitude of the SFQ pulse which results in a pulse with a rise time near the practical maximum for the 0.2~\mAumsq JJ process. We maximize the SFQ pulse amplitude by increasing the JJ $I_c$ as this also increases the characteristic voltage and thus the peak voltage of the output SFQ pulse. Pulse-area-quantization means this also shortens the total duration of the SFQ pulse. This is expected to be beneficial in both maximizing the rise time \cite{askerzade2006josephson} and, in the case such an impulse generator is used to supply the sampling strobe, providing an even larger maximum amplitude pulse than is possible using a Faris pulser. For comparison, our simulated Faris pulser strobe delivers, through the 2~$\Omega$ strobe coupling resistor, a peak current pulse of 400~$\mu$A while the DC/SFQ converter achieves a 600~$\mu$A pulse.

As in the Faris pulser DUT case, we simulated sampler reconstruction of the target signal as a function of the comparator $\beta_C$ via tuning $R_{sh}$. Fig.~\ref{fig:sfq_pulse_vs_Rshunt} shows the simulated circuit and simulation results, which echo the findings of Fig.~\ref{fig:faris_pulser_pulse_vs_Rshunt}. Again, a notable observation is that the sampler is expected to fail to reconstruct the pulser DUT amplitude but should be relatively faithful in measuring the signal FWHM. Finally, we note the similar strong dependence on distortion for sampling finite-duration signals when the comparator is appreciably underdamped ($\beta_C \gtrsim 3$).

Different device variants using all permutations of Faris pulsers and DC/SFQ converters as both the strobe and impulse DUT, designed according to Fig.~\ref{fig:sfq_pulse_vs_Rshunt}(a), were fabricated on the same wafer as the sampler and DUTs presented in the rest of the text. Measuring the chip variant with a Faris pulser strobe and DC/SFQ DUT, we successfully sample the output of a DC/SFQ converter while operating at 3.6~K. Rather than apply a dc flux offset to the DUT trigger we realize a programmable, tunable strobe-signal delay using a motor-controlled mechanical delay line installed on the strobe trigger. This delay line has a maximum delay of  170 ps and is equipped with a variable follower resistor whose resistance is proportional to the delay.

To minimize hysteresis and position jitter the motor is first driven to its maximum and then minimum delay, after which we begin the sampling measurement. For each strobe-signal trigger delay we step the delay line motor by its minimum increment and only in the forward (increasing strobe delay) direction. The follower resistance is measured to determine the programmed delay at each motor position. The data in Fig.~\ref{fig:sfq_pulse_vs_Rshunt}(e) were acquired in this manner over three separate scans. Reverse scans were also explored with no observable difference in sampler output. The repeatability of these individual scans is surprisingly consistent but it is clear the minimum motor step is insufficient for sampler measurements targeting a minimum relative delay spacing of $\lesssim 0.1$~ps. We are investigating other methods for sweeping delay to provide 0.1 ps or smaller delay intervals over a wide ($>$~100~ps) time sweep. The raw sampled pulse data included a small (0~$\mu$A--7~$\mu$A), slowly-varying background signal due to on-chip ground bounce and capacitive feedthrough of the trigger signal. This was subtracted from the data presented in Fig.~\ref{fig:sfq_pulse_vs_Rshunt}(e). To measure this spurious signal the DC/SFQ DUT was sampled both with its dc bias current on (SFQ pulse is output) and with it set to zero (no SFQ pulse is output). For both cases the same input square wave (125~ps period, 50\% duty cycle) trigger signal amplitude was provided.

\begin{figure*}
    \centering
    \includegraphics[width=0.95 \textwidth]{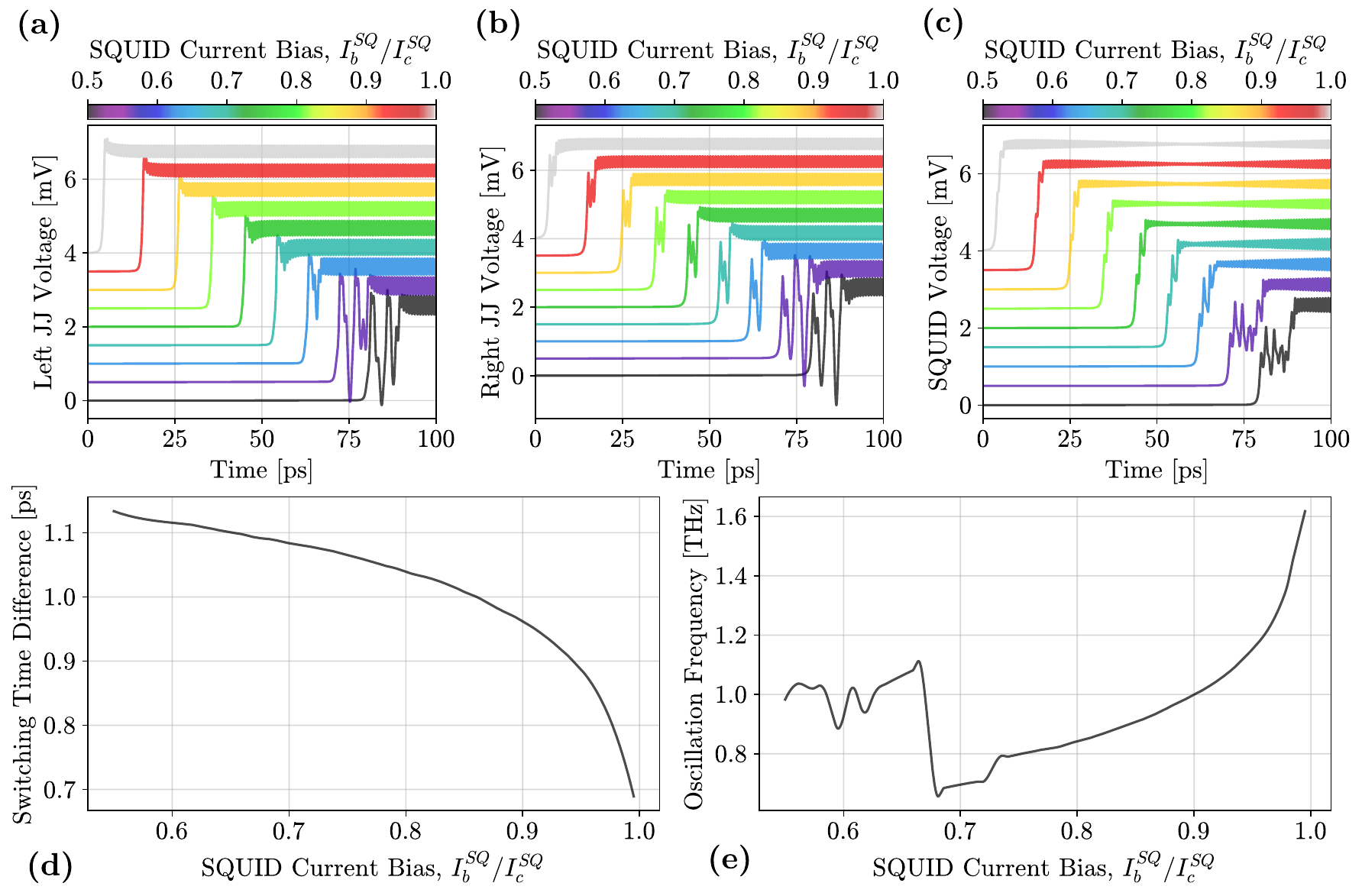}
    \caption{Simulation of the latching dynamics of the SQUID DUT alone as a function of the DUT bias current $I_b^{SQ} / I_c^{SQ}$. \textbf{(a)} Voltage of the lefthand SQUID JJ at selected current biases. \textbf{(b)} Voltage of the righthand SQUID JJ. \textbf{(c)} Total SQUID output voltage. At low current biases the large swings in voltage on the rising edge of the step signal the SQUID generates are due to the right JJ switching and driving currents (at its Josephson frequency) through the left JJ before it switches. Eventually, as the SQUID bias is increased, the time delay between the right and left JJs switching is reduced sufficiently for the right JJ's Josephson oscillations to only impact the left JJ voltage, and thus the full SQUID, while the left JJ is on it's rising edge. \textbf{(d)} Delay between the switching time of the right and left JJs. Switching time is defined as when the JJ voltage exceeds 10\% of the gap voltage. \textbf{(e)} Extracted oscillation frequency on the rising edge of the SQUID signal obtained via a polynomial plus sinusoidal fit as in Sec.~\ref{sec:dut_Ib_sweep}.}
    \label{fig:squid_latching_dynamics_sim}
\end{figure*}

\section{\label{app:squid_latching_dynamics}SQUID Latching Dynamics}
To verify the oscillatory features on the rising edge of the latching SQUID signal observed in Sec.~\ref{sec:dut_Ib_sweep} we perform simulations, as a function of current bias, of the SQUID DUT alone with its 12~$\Omega$ coupling resistor shorted to ground. The SQUID is symmetric with each JJ $I_c = 1$~mA and the flux trigger is a step function pulse with a rise time of 100~ps. Using this simulation we may track each junction's voltage independently, as well as the total SQUID voltage. These results are shown in Fig.~\ref{fig:squid_latching_dynamics_sim}, clearly demonstrating that oscillations on the rising edge of the SQUID signal are expected. Furthermore, we also show that these oscillations arise from the delay in switching time -- defined here as the time at which the JJ voltage exceeds 10\% of the gap voltage of 2.8~mV -- between the left and right JJ. Indeed, with the simulated flux coupling inductor polarity and the direction of the screening current, the right JJ switches as much as 1.1~ps before the left JJ switches (Fig.~\ref{fig:squid_latching_dynamics_sim}(d)) and this delay decreases as $I_b^{SQ}$ is increased. As a final check we again perform a polynomial and sinusoidal fit to the rising edge of the simulated SQUID voltage, this time to the \mbox{50\%--90\%} voltage, and obtain the fitted oscillation frequency shown in Fig.~\ref{fig:squid_latching_dynamics_sim}(e); which also display a distinct transition in frequency. We choose the \mbox{50\%--90\%} portion of the SQUID step function in simulation and the $\geq 90\%$ portion in measurement to more accurately target the SQUID JJ beat frequency oscillations. In measurement (Fig.~\ref{fig:squid_Ib_sweep}) these oscillations arise in the central portion of the step and persist at the same frequency up to the very top of the step. For the simulations the beat frequency of interest also appears in the step rising edge, however, once the SQUID has reached it's asymptotic maximum each JJ is undergoing persistent Josephson oscillations at a frequency above the sampler bandwidth and this masks the beat oscillation. Finally, it is difficult to fit a polynomial to a sharp step function without a large number of terms and this is the reason for choosing only a portion of the signal for performing the fit.

\begin{figure}
    \centering
    \includegraphics[width=0.5\textwidth]{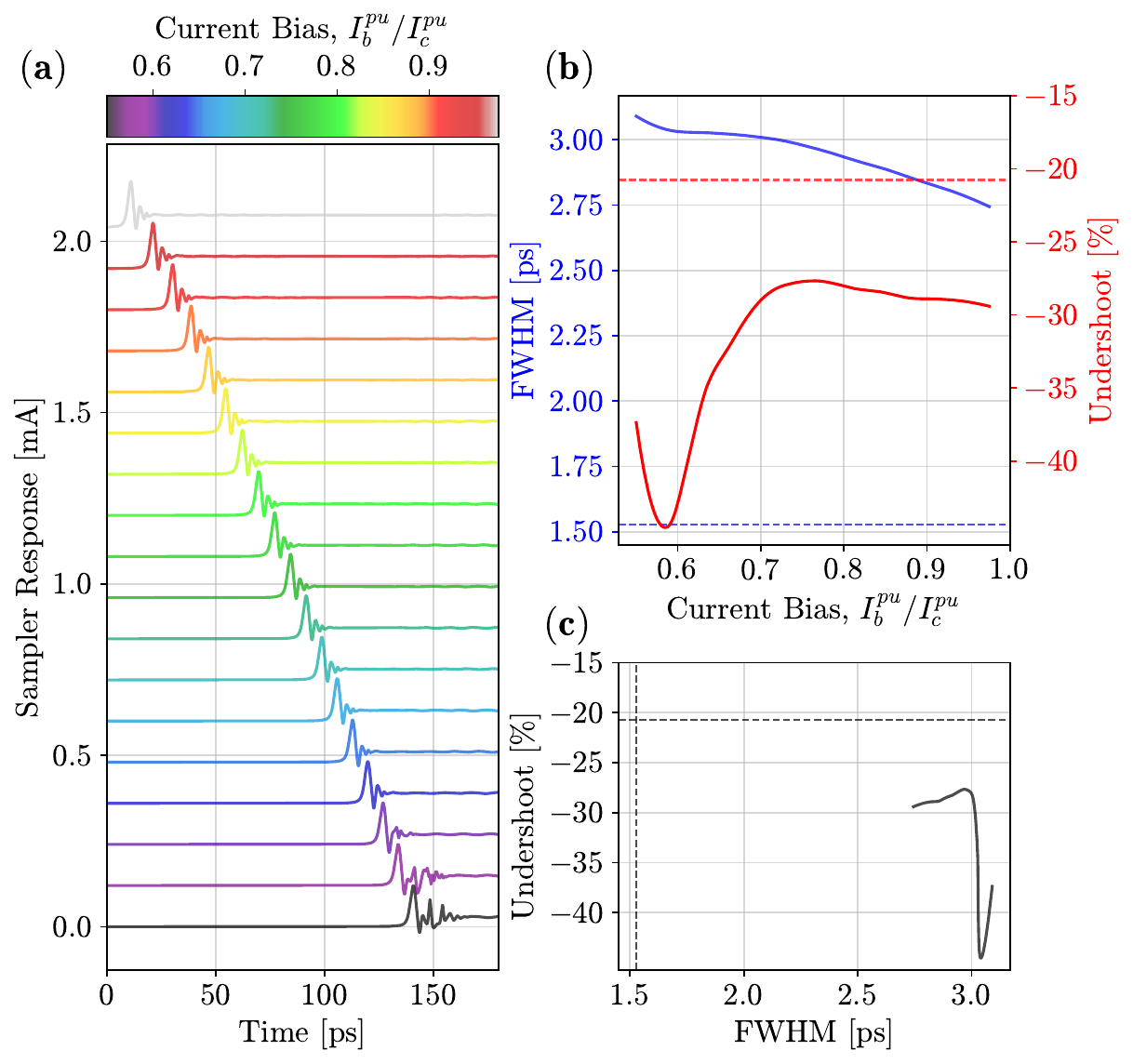}
    \caption{Simulated sampler output of the Faris pulser DUT as a function of the pulser DUT current bias. \textbf{(a)} Sampled waveforms offset for clarity. \textbf{(b)} Extracted sampled pulse FWHM and distortion (undershoot relative to the peak pulse height). \textbf{(c)} Dependence of the undershoot as a function of the sampled FWHM.}
    \label{fig:pulser_Ib_sweep_sim}
\end{figure}

\section{\label{app:sampling_faris_pulser}Simulations of Sampling Faris Pulser DUT}
Our simulation framework is also capable of performing full sampling simulations as a function of DUT bias. This allows us to compare the experimental results of Fig.~\ref{fig:squid_Ib_sweep} and Fig.~\ref{fig:pulser_Ib_sweep} to simulation. As an example we perform a Faris pulser DUT current bias sweep sampler simulation to track the evolution of the sampled waveform. The strobe bias is fixed at $I_b^{str} / I_c^{str} = 0.9$. Fig.~\ref{fig:pulser_Ib_sweep_sim} shows the results of these simulations and we observe high quality agreement with the measurements of Fig.~\ref{fig:pulser_Ib_sweep}. Additionally there is correspondence with the SQUID dynamics simulations detailed in Fig.~\ref{fig:squid_latching_dynamics_sim} where at low bias we see multiple complex post-pulse oscillations due to interference in the voltages sources from the two JJs in the Faris pulser's SQUID. These oscillations become coherent at high bias and are still visible in simulation even at high DUT bias. In experiment we cannot resolve these features once the bias exceeds $I_b^{pu} / I_c^{pu} \sim 0.9$ and we attribute this to the smoothing effects arising from the comparator dynamics convolution with the DUT signal.

\end{document}